\documentclass[11pt,a4paper]{article}
\pdfoutput=1
\usepackage{jcappub}
\usepackage{rotating}
\usepackage{array}
\usepackage{amsmath}
\usepackage[normalem]{ulem}
\usepackage{slashed}
\usepackage{booktabs}
\usepackage[pdftex,table]{xcolor}
\usepackage{units}
\usepackage{multirow}

\arxivnumber{DESY-18-028, TTK-18-08}

\title{Model-independent comparison of annual modulation and total rate with direct detection experiments}

\author[a]{Felix Kahlhoefer,}
\author[b,c]{Florian Reindl,}
\author[d]{Karoline Sch\"{a}ffner,}
\author[e]{\\Kai Schmidt-Hoberg}
\author[e]{and Sebastian Wild}

\affiliation[a]{Institute for Theoretical Particle Physics and Cosmology (TTK), RWTH Aachen University, \\ D-52056 Aachen, Germany}
\affiliation[b]{Institut f\"{u}r Hochenergiephysik der \"{O}AW, A-1050 Wien, Austria}
\affiliation[c]{Atominstitut, Technical University Vienna, A-1020 Wien, Austria}
\affiliation[d]{Gran Sasso Science Institute (GSSI), I-67100 L'Aquila, Italy} 
\affiliation[e]{Deutsches Elektronen-Synchrotron (DESY), Notkestrasse 85, D-22603 Hamburg, Germany}

\emailAdd{kahlhoefer@physik.rwth-aachen.de}
\emailAdd{florian.reindl@oeaw.ac.at}
\emailAdd{karoline.schaeffner@lngs.infn.it}
\emailAdd{kai.schmidt-hoberg@desy.de}
\emailAdd{sebastian.wild@desy.de}

\abstract{
  The relative sensitivity of different direct detection experiments depends sensitively on the astrophysical distribution and particle physics nature of dark matter, prohibiting a model-independent comparison. The situation changes fundamentally if two experiments employ the same target material. We show that in this case one can compare measurements of an annual modulation and exclusion bounds on the total rate while making no assumptions on astrophysics and no (or only very general) assumptions on particle physics. In particular, we show that the dark matter interpretation of the DAMA/LIBRA signal can be conclusively tested with COSINUS, a future experiment employing the same target material. We find that if COSINUS excludes a dark matter scattering rate of about $0.01\,\text{kg}^{-1}\,\text{days}^{-1}$ with an energy threshold of $1.8\,$keV and resolution of $0.2\,$keV, it will rule out all explanations of DAMA/LIBRA in terms of dark matter scattering off sodium and/or iodine.}


\begin{document}
\maketitle
\flushbottom

\section{Introduction}
\label{sec:introduction}

When lamenting the absence of conclusive evidence for particle dark matter (or in fact any kind of new physics beyond the Standard Model), it is often forgotten that we have been observing for many years a statistically highly significant signal compatible with predictions for the scattering of dark matter (DM) particles from the Galactic halo. The combined data from the DAMA/NaI and the DAMA/LIBRA experiments show evidence for an annual modulation of the event rate at the 9.3$\sigma$ level~\cite{Bernabei:2013xsa}, which agrees with the DM hypothesis in both period and phase.\footnote{One of the few unexpected features in the DAMA data is a slight time-dependence of the modulation amplitude~\cite{Kelso:2017zwr}.} Even though the NaI detectors employed by the DAMA collaboration do not allow to distinguish nuclear and electron recoils, it has not been possible to identify a plausible background that could account for the observed signal while satisfying all cross-checks performed by the DAMA collaboration (see refs.~\cite{Blum:2011jf,Pradler:2012qt,Pradler:2012bf,Bernabei:2012wp,Davis:2014cja,Bernabei:2014tqa,Klinger:2015vga} for recent discussions).

The great trouble with the DAMA signal is its incompatibility with exclusion limits obtained from other direct detection experiments. For the simplest assumptions, i.e.\ spin-independent or spin-dependent interactions between DM and nuclei, a number of experiments exclude the parameter region preferred by DAMA by several orders of magnitude~\cite{Angloher:2014myn,Akerib:2016vxi,Hehn:2016nll,Cui:2017nnn,Aprile:2017iyp,Amole:2017dex,Agnese:2017njq,Agnese:2017jvy,Agnes:2018fwg,Agnes:2018ves}. But even more exotic hypotheses, such as leptophilic DM~\cite{Kopp:2009et}, inelastic DM~\cite{TuckerSmith:2001hy} or isospin-violating DM~\cite{Feng:2011vu}, are strongly disfavoured by recent searches (see e.g.\ ref.~\cite{Aprile:2017yea}).

While it is very challenging to construct DM models that bring all experimental results into agreement (see refs.~\cite{Dolan:2014ska,DelNobile:2015lxa,Scopel:2015eoh,Scopel:2015baa,Catena:2016hoj,Kang:2018dlc} for some recent attempts), it is even more difficult to generally exclude a DM explanation of the DAMA signal. The reason is that any comparison between different direct detection experiments requires a number of assumptions on both the astrophysical distribution and the particle physics properties of DM and the conclusions may depend sensitively on these assumptions. A number of halo-independent methods have been developed to deal with the former issue, allowing for a statistically meaningful comparison of different experiments without the need to assume a specific DM density or velocity distribution~\cite{Fox:2010bz,McCabe:2011sr,Frandsen:2011gi,DelNobile:2013cva,Ibarra:2017mzt,Gondolo:2017jro,Catena:2018ywo}.

The best way to reduce the need to make specific assumptions on the particle physics properties of DM would be an independent test of the DAMA result with NaI-based detectors. Indeed, a number of experimental collaborations are currently pursuing this strategy. Most of these, namely SABRE~\cite{Shields:2015wka}, ANAIS~\cite{Amare:2014jta}, PICO-LON~\cite{Fushimi:2015sew} and COSINE~\cite{Thompson:2017yvq} (the latter being a joint effort of the DM-Ice~\cite{Cherwinka:2014xta} and KIMS~\cite{Kim:2012rza} collaborations), aim to search for an annual modulation in their event rates in much the same way as the original DAMA experiment. The COSINUS collaboration~\cite{Angloher:2016ooq}, in contrast, has the more ambitious goal to develop a cryogenic NaI detector that can distinguish nuclear and electron recoils. This approach will allow to significantly suppress background levels, making it possible for the first time to test the DAMA signal using the absolute rate of nuclear recoils in NaI detectors. At the same time, the possibility to measure phonon signals will allow COSINUS to significantly lower the nuclear recoil energy threshold compared to DAMA, further increasing its sensitivity.

Nevertheless, the comparison of a total rate and a modulation amplitude appears to require specific assumptions on the DM velocity distribution, which determines the modulation fraction of the signal. Existing halo-independent methods deal with this issue by making very conservative assumptions on the modulation fraction, typically allowing modulation fractions up to 100\%.\footnote{A notable exception are refs.~\cite{HerreroGarcia:2011aa,HerreroGarcia:2012fu,Bozorgnia:2013hsa,Blennow:2015gta,Herrero-Garcia:2015kga}, which derive halo-independent constraints on the modulation amplitude from constraints on the total rate. While very similar in spirit, the difference between these works and ours is that we explicitly construct the velocity distribution that maximizes the modulation amplitude rather than deriving conservative upper bounds.} In the present work we show that for the case of several experiments employing the same detector material, significantly stronger consistency requirements can be made, leading to potentially much tighter constraints. Moreover, we will show that halo-independent methods can be generalised to consider not only arbitrary DM velocity distributions but also very general classes of interactions between DM and nuclei. This approach is therefore ideally suited for comparing future results from COSINUS with the DAMA signal.

\begin{figure}
\centering
\hspace*{-0.7cm}
\includegraphics[scale=0.9]{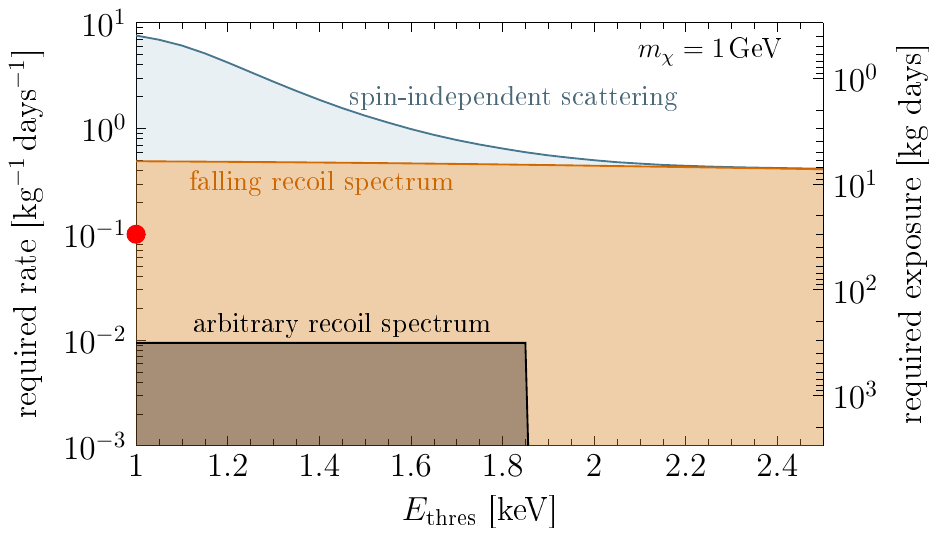}\hspace*{0.2cm}
\includegraphics[scale=0.9]{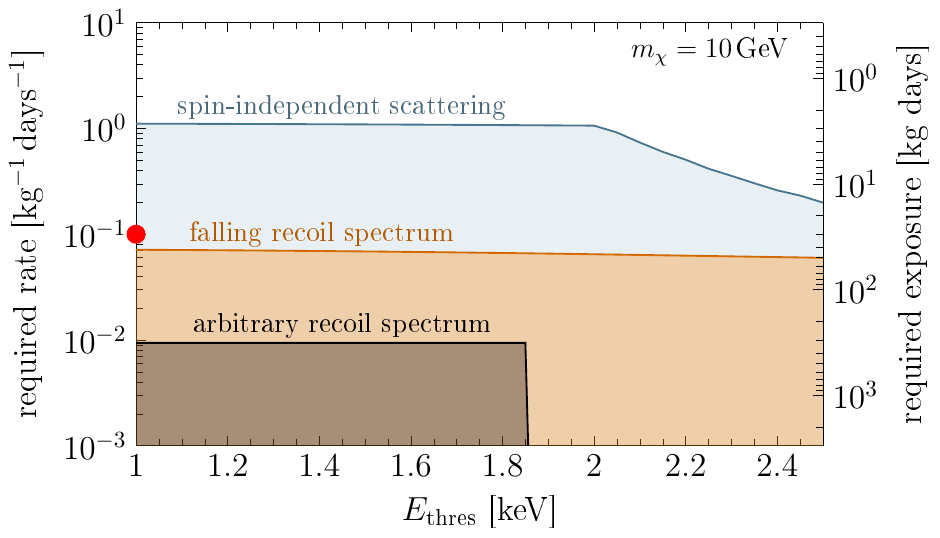}\\[0.2cm]
\hspace*{-0.7cm}
\includegraphics[scale=0.9]{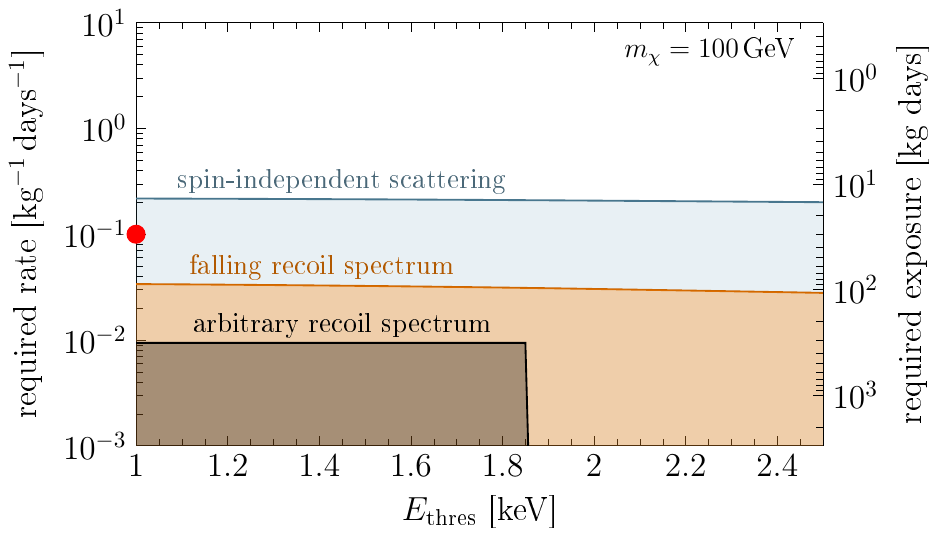}\hspace*{0.2cm}
\includegraphics[scale=0.9]{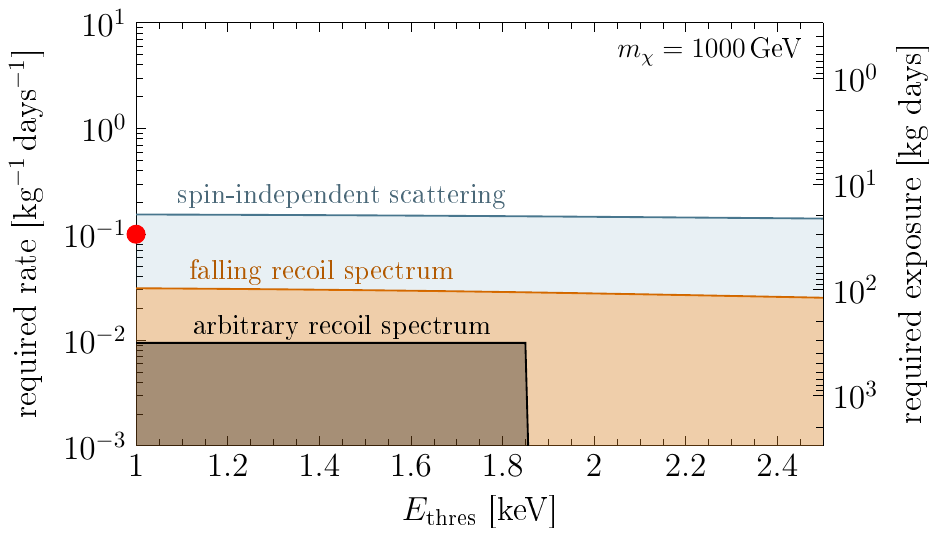}
\caption{\small COSINUS exclusion power, defined as the bound on the total rate (or equivalently the total exposure with zero observed events) that COSINUS must achieve for excluding DAMA in a halo-independent way, as a function of the assumed threshold in COSINUS for different DM masses. More general assumptions correspond to weaker exclusion power, meaning that stronger bounds are necessary to achieve an exclusion. The most general bound, valid for arbitrary DM-nucleus scattering, is derived in section~\ref{sec:efficiencies}. The assumption of falling recoil spectra, which is more restrictive but still covers many cases of interest, is discussed in section~\ref{sec:classes}. The specific example of halo-independent constraints on spin-independent scattering is considered in section~\ref{sec:specific}. The red dot in each panel indicates the design sensitivity of COSINUS, $E_\text{thres} = 1\,\text{keV}$ and $R_\text{COSINUS}^\text{bound} = 0.1\,\mathrm{kg^{-1}\,days^{-1}}$.}
\label{fig:1}
\end{figure}

A particular emphasis is placed on the question how to optimise the design of the COSINUS detector. As in all direct detection experiments, the detector development faces two competing goals: lowering the energy threshold and improving the bound on the event rate above threshold (e.g.\ by increasing the detector mass). We show that for a halo-independent test of the DAMA signal, it is highly desirable to achieve an energy threshold of about $1.8\,\mathrm{keV}$ (based on conservative assumptions on the quenching factors in DAMA and the energy resolution in COSINUS), while a further reduction of the threshold is less advantageous than e.g.\ an increase in exposure. These conclusions are illustrated in figure~\ref{fig:1}, which shows the bound on the total event rate that COSINUS must achieve in order to exclude the DAMA signal in a halo-independent way as a function of the COSINUS threshold. The different lines correspond to different assumptions on the interactions of DM, which will be explained in the following chapters.

This paper is structured as follows: In section~\ref{sec:DAMA} we review the DAMA experiment and the COSINUS proposal, and summarise the assumptions made for our specific implementation of these experiments. In section~\ref{sec:efficiencies} we present the simplest~-- but also most conservative~-- way to compare the two experiments by comparing the respective efficiencies. Section~\ref{sec:classes} discusses a more constraining approach, valid for rather general classes of recoil spectra. We compare the results obtained from this approach with the ones obtained for more specific assumptions on the interactions of DM in section~\ref{sec:specific}. Our conclusions are presented in section~\ref{sec:conclusions}, while additional technical details can be found in appendix~\ref{app:proof}.

\section{Direct detection with DAMA and COSINUS}
\label{sec:DAMA}

To introduce the relevant notation and conventions, let us briefly review the formalism for DM direct detection. The differential event rate with respect to recoil energy is given by
\begin{equation}
\frac{\text{d}R^\text{T}}{\text{d}E_\text{R}} = \frac{\xi_\mathrm{T} \rho}{m_\mathrm{T} \, m_\chi}  \int_{v_\text{min}}^\infty v f(\boldsymbol{v} + \boldsymbol{v}_\text{E}(t)) \frac{\text{d} \sigma^\text{T}}{\mathrm{d} E_{\text{R}}}  \text{d}^3 v\; ,
\label{eq:dRdE}
\end{equation}
where $m_\chi$ and $m_\mathrm{T}$ are the mass of the DM particle and the target nucleus T, $\xi_\text{T}$ is the mass fraction of T relative to the total detector mass,  $\rho$ and $f(\mathbf{v})$ are the local DM density and velocity distribution in the rest frame of the Sun, $\boldsymbol{v}_\text{E}(t)$ is the velocity of the Earth and $v=|\boldsymbol{v}|$.\footnote{It is conventional to define both $f(\mathbf{v})$ and $\boldsymbol{v}_\text{E}(t)$ in the Galactic rest frame, where the velocity distribution often takes a very simple form. However, in the context of halo-independent methods it will be more convenient to specify $f(\mathbf{v})$ in the rest frame of the Sun.} In order to transfer an energy $E_\text{R}$ in an elastic collision, the incoming DM particle must have a \emph{minimum velocity} of
\begin{equation}\label{eq:vmin}
v_\text{min}(E_\text{R}) = \sqrt{\frac{m_\mathrm{T} E_\text{R}}{2 \, \mu_{\mathrm{T}\chi}^2}} \;,
\end{equation}
where $\mu_{\mathrm{T}\chi} = m_\mathrm{T} \, m_\chi / (m_\mathrm{T} + m_\chi)$ is the reduced mass of the DM-nucleus system.

For reasons that will become apparent below, we will focus on cut-and-count experiments considering a single signal region in energy, i.e.\ disregarding the spectral information of the signal. Such an experiment can be fully characterised by the total exposure $\mathcal{E}$ and the efficiency functions $\epsilon^\text{T}(E_\text{R})$ that quantify the probability for a nuclear recoil of the target nucleus T with energy $E_\text{R}$ to lead to an event in the signal region. In terms of these quantities, the number of predicted events is given by
\begin{equation}
 N = \mathcal{E} \sum_\text{T} \int \frac{\text{d}R^\text{T}}{\text{d}E_\text{R}} \, \epsilon^\text{T}(E_\text{R}) \, \text{d}E_\text{R} \; .
\end{equation}
The total event rate in the signal region is given by
\begin{equation}
 R = \frac{N}{\mathcal{E}} =  \sum_\text{T} \int \frac{\text{d}R^\text{T}}{\text{d}E_\text{R}} \, \epsilon^\text{T}(E_\text{R}) \, \text{d}E_\text{R} \; .
 \label{eq:rate}
\end{equation}
A null result allows an experiment to place an upper bound on $N$ for a given exposure or, equivalently, an upper bound on $R$. For example, in the case of zero observed events, it is possible to exclude $N > 2.71$ at 90\% confidence level (CL), leading to the upper bound $R < 2.71 / \mathcal{E}$.

In general, both $N$ and $R$ depend on time because of the time-dependence of $\mathbf{v}_\text{E}(t)$. If the DM velocity distribution is isotropic in the Galactic rest frame, one expects an approximately sinusoidal annual modulation that peaks around the 1st of June~\cite{Drukier:1986tm,Freese:2012xd}. This prediction is in good agreement with the time-dependence of the DAMA signal. We therefore define the mean rate\footnote{Note that for approximately sinusoidal modulation with period of one year, our definition of $\bar{R}$ agrees with the time-averaged rate $\langle R \rangle = \tfrac{1}{t} \int R(t') \, \mathrm{d}t'$.}
\begin{align}
\bar{R} = \frac12 \left( R(t = \text{1st of June}) + R(t = \text{1st of December}) \right) \; ,
\label{eq:Rbar_definition}
\end{align}
and the modulation amplitude
\begin{align}
S = \frac12 \left( R(t = \text{1st of June})  - R(t = \text{1st of December}) \right) \; .
\label{eq:S_definition}
\end{align}
The results of the DAMA experiment can then be summarised in terms of the modulation amplitude observed in different energy bins. The greatest evidence for an annual modulation is seen in the energy range $E^\text{obs}_\mathrm{R} \in \left[ 2.5\,\mathrm{keVee}, 3.5\,\mathrm{keVee}\right]$, where $E^\text{obs}_\mathrm{R}$ denotes observed recoil energy after quenching (see below). In this energy range the combined data from DAMA/NaI and DAMA/LIBRA imply~\cite{Bernabei:2013xsa}
\begin{equation}
S = (2.34 \pm 0.28)\cdot 10^{-2} \, \mathrm{kg^{-1}\,days^{-1}} \; .
\label{eq:SDAMA}
\end{equation}
In the following we will focus on this energy range and study whether COSINUS can test the hypothesis that the signal observed there is due to interactions between DM and nuclei. If COSINUS can exclude this hypothesis, it will be impossible to find a DM model that provides a satisfactory explanation of the DAMA signal. We have checked explicitly that our conclusions do not depend sensitively on the energy range that we consider.

In order to predict the modulation amplitude in the DAMA signal region for a given differential event rate, one must first of all convert the physical recoil energy $E_\mathrm{R}$ into a quenched energy $E'_\mathrm{R} = Q \, E_\mathrm{R}$, where $Q$ denotes the quenching factor.\footnote{To distinguish between physical recoil energy and reconstructed (quenched) recoil energy, we use the unit keVee (ee = electron equivalent) when referring to the latter.} For the present analysis we adopt the values $Q = 0.3$ and 0.09 for sodium and iodine scatters, respectively~\cite{Bernabei:1996vj}. Some recent measurements indicate even smaller quenching factors~\cite{Xu:2015wha}, which would further strengthen the sensitivity of COSINUS with respect to the DAMA data as the COSINUS detection efficiency is a monotonically increasing function of energy. To calculate the efficiency functions $\epsilon^\text{T}(E_\mathrm{R})$, we follow refs.~\cite{Savage:2008er,Bernabei:2008yh} and assume a Gaussian energy resolution with $\sigma(E'_\mathrm{R}) = (0.448\,\text{keVee}) \sqrt{E'_\mathrm{R}/\text{keVee}} + 0.0091 E'_\mathrm{R}$. We only consider fluctuations of up to four standard deviations, in order to avoid an unphysical extrapolation.

\bigskip

In contrast to DAMA, COSINUS aims to use NaI within a {\it cryogenic detector}, simultaneously acquiring a phonon (heat) and a light signal from a particle interaction. The phonon channel yields a precise and unquenched, thus, particle-independent measurement of the energy deposited in the crystal; the scintillation light provides particle discrimination on an event-by-event basis.  This technology has been shown to be extraordinarily sensitive and
allowed the CRESST collaboration to lead the field of direct dark matter detection on the low DM mass frontier~\cite{Petricca:2017zdp}.
In the past two years COSINUS successfully proved for the first time that cryogenic operation of NaI is possible in ref.~\cite{Angloher:2017sft}, leading to the commissioning of the final detector design~\cite{Reindl:2017bun}.\footnote{In parallel to detector optimisation the COSINUS collaboration is also preparing measurements of the quenching factors for scatterings off sodium and iodine at cryogenic temperatures by irradiating a COSINUS detector to neutrons from AmBe sources and/or a neutron beam~\cite{Strauss:2014zia}.} The light channel already exceeds the final performance goals while further improvements in the phonon channel are necessary in order to further decrease the threshold. 

Our implementation of the COSINUS experiment largely follows ref.~\cite{Angloher:2016ooq}. We assume that physical recoil energy $E_\mathrm{R}$ and observed recoil energy $E_\mathrm{R}^\text{obs}$ are related via a Gaussian energy resolution with $\sigma_{E} = 0.2\,\,\mathrm{keV}$, where once again we only take into account fluctuations of up to $4 \sigma_E$. The probability for detecting a nuclear recoil with observed energy $E_\mathrm{R}^\text{obs}$ is given by 
\begin{align}
\epsilon_\text{det}(E_\mathrm{R}^\text{obs}) = \Theta(E_\mathrm{R}^\text{obs} - E_\text{thres}) \times \begin{cases} 0 &\mbox{for } E_\mathrm{R}^\text{obs} < 1\,\text{keV} \,,\\
0.3 \cdot \frac{E_\mathrm{R}^\text{obs}}{\text{keV}} - 0.1 & \mbox{for } 1\,\text{keV} \leq E_\mathrm{R}^\text{obs} \leq 2\,\text{keV} \,,\\
0.5  &\mbox{for } E_\mathrm{R}^\text{obs} > 2 \,\text{keV} \,,
\end{cases}
\end{align}
where $\Theta(x)$ is the Heaviside step function and $E_\text{thres}$ denotes the low-energy threshold of the signal region considered by COSINUS. Unless explicitly stated otherwise, we adopt the value $E_\text{thres} = 1\,\text{keV}$.

A crucial part of the COSINUS strategy to reduce backgrounds is to reject events with large light yield, as these are much more likely to arise from electron recoils than from nuclear recoils. The signal region is therefore restricted to events with a light yield below the mean of the sodium nuclear recoil band (see figure~2 in ref.~\cite{Angloher:2016ooq}). The final efficiency function is then given by $\epsilon(E_\mathrm{R}^\text{obs}) = \epsilon_\text{det}(E_\mathrm{R}^\text{obs}) \times \epsilon_\text{ly}(E_\mathrm{R}^\text{obs})$, where the latter factor quantifies the probability that an event with reconstructed energy $E_\mathrm{R}^\text{obs}$ has sufficiently low light yield to fall into the signal region. For sodium recoils, one has by construction $\epsilon_\text{ly} = 0.5$, whereas a fit of the iodine nuclear recoil band gives
\begin{align}
\epsilon_\text{ly}(E_\mathrm{R}^\text{obs}) \approx 1 - 0.42 \exp(-0.89 E_\mathrm{R}^\text{obs}/\text{keV})
\end{align}
for iodine recoils.

It is important to emphasise that the ability of the COSINUS detector to discriminate between electron recoils and nuclear recoils is reduced at low energies. In other words, a lowering of $E_\text{thres}$ may lead to an increase in the number of observed background events and hence to a weaker bound on $R$. One of the central questions of the subsequent sections will be to determine under which conditions it will be more desirable for COSINUS to lower $E_\text{thres}$ or to strive for a more stringent bound on $R$.

\bigskip

\begin{figure}
\centering
\hspace*{-1.2cm}
\includegraphics[width=0.6\textwidth]{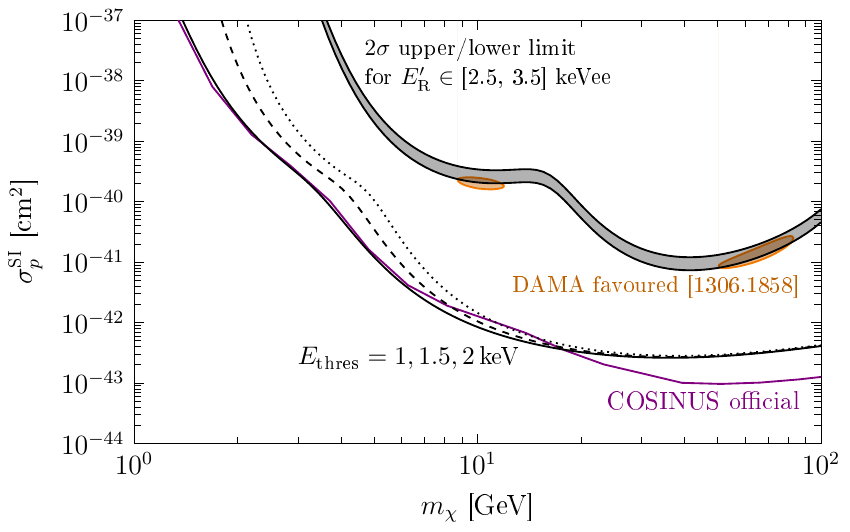}
\caption{\small Comparison of the projected sensitivity of COSINUS (black lines) and the parameter regions favoured by DAMA (shaded grey) for spin-independent interactions and a Maxwell-Boltzmann velocity distribution. The three different lines for COSINUS correspond to different assumptions on $E_\text{thres}$. For comparison we show the COSINUS expected sensitivity for $E_\text{thres} = 1\,\text{keV}$ derived in ref.~\cite{Angloher:2016ooq} using an optimum interval analysis (violet line). The orange shaded regions, which are taken from fig.~2 of ref.~\cite{Kelso:2013gda}, indicate the parameter regions favoured by DAMA when including 8 separate bins (90\% CL, rescaled to $\rho = 0.3\,\text{GeV}\,\text{cm}^{-3}$).}
\label{fig:standard}
\end{figure}

To conclude this section, we compare in figure~\ref{fig:standard} the projected sensitivity of the COSINUS experiment with the parameter region favoured by DAMA for standard assumptions on the properties of DM~\cite{Lewin:1995rx}. Specifically we assume spin-independent interactions with
\begin{equation}
\frac{\text{d}\sigma^\text{T}}{\text{d} E_\text{R}} = A^2 \, F^2(E_{\text{R}}) \frac{m_\mathrm{T} \, \sigma_p}{2 \, \mu_{p\chi}^2 \, v^2}\;,
\label{eq:SI}
\end{equation}
where $A$ and $F(E_\mathrm{R})$ denote the mass number and form factor of the target nucleus, and $\sigma_p$ and $\mu_{p\chi}$ denote the DM-proton scattering cross section and reduced mass, respectively. We set $\rho = 0.3\,\mathrm{GeV \, cm^{-3}}$ and adopt a Maxwell-Boltzmann distribution with 
most probable speed $v_0 = 220\,\text{km}\,\text{s}^{-1}$ and a cut-off at $v_\text{esc} = 544\,\text{km}\,\text{s}^{-1}$ for $f(v)$ in the Galactic rest frame. To obtain the corresponding velocity distribution in the rest frame of the Sun, we take the velocity of the Sun in the Galactic rest frame to be $v_\text{sun} = 230\,\mathrm{km\,s^{-1}}$.

We assume that COSINUS can set an upper bound of $R < 0.1 \, \mathrm{kg^{-1} \, days^{-1}}$ (corresponding e.g.\ to a 95\% CL upper limit for an exposure of $105\,\mathrm{kg \, days}$ and 5 observed events) and illustrate the impact of changing the low-energy threshold by considering three different values of $E_\text{thres}$. Our estimated sensitivity for $E_\text{thres} = 1\,$keV agrees quite well with the one from ref.~\cite{Angloher:2016ooq}, which makes use of the optimum interval method~\cite{Yellin:2002xd} to derive constraints. Since we implement only a single energy range for DAMA, we do not find two separate signal regions (corresponding respectively to scattering dominantly off sodium and dominantly off iodine) but instead obtain a band of allowed parameters. In particular, our implementation of DAMA is conservative, because it leads to a significantly larger allowed parameter region.

The central conclusion from figure~\ref{fig:standard} is that for the assumptions of spin-independent interactions and a Maxwell-Boltzmann velocity distribution COSINUS should be able to exclude the DAMA region by about two orders of magnitude in cross section. However, as pointed out in section~\ref{sec:introduction}, the DAMA hypothesis is already robustly excluded for standard assumptions by a number of experiments. We will therefore now turn to the much more interesting question how the sensitivity of COSINUS compares to DAMA for less restrictive assumptions.

\section{Halo- and model-independent comparison}
\label{sec:efficiencies}
We begin by discussing the most general~-- and hence most conservative~-- way of comparing DAMA and COSINUS. This approach is based on the very simple observation that the modulation amplitude in a given experiment cannot exceed the mean rate~\cite{Frandsen:2011gi}:
\begin{equation}
 S \leq \bar{R} \; ,
 \label{eq:SleqR}
\end{equation}
which directly follows from eqs.~(\ref{eq:Rbar_definition}) and~(\ref{eq:S_definition}). We now define $S_\text{DAMA}^\text{bound} \equiv 1.78 \cdot 10^{-2} \, \mathrm{kg^{-1}\,days^{-1}}$ as the smallest value of the modulation amplitude compatible with the DAMA measurement at the $95\%$ CL, see eq.~(\ref{eq:SDAMA}),  and $R_\text{COSINUS}^\text{bound}$ as the 95\% CL upper bound on the mean total event rate in COSINUS. Using eq.~(\ref{eq:SleqR}), the modulation amplitude $S_\text{DAMA}$ in DAMA must then satisfy
\begin{align}
S_\text{DAMA}^\text{bound} \leq S_\text{DAMA} \leq \bar{R}_\text{DAMA} = \sum_\text{T} \int \frac{\text{d}\bar{R}^\text{T}}{\text{d}E_\text{R}} \, \epsilon_\text{DAMA}^\text{T}(E_\text{R}) \, \text{d}E_\text{R} \,,
\end{align}
where we define $\text{d}\bar{R}^\text{T}/\text{d}E_\text{R}$ analogously to eq.~(\ref{eq:Rbar_definition}) as the mean differential recoil rate during the year. On the other hand, the upper bound from COSINUS implies
\begin{align}
R_\text{COSINUS}^\text{bound} \geq \bar{R}_\text{COSINUS} = \sum_\text{T} \int \frac{\text{d}\bar{R}^\text{T}}{\text{d}E_\text{R}} \, \epsilon_\text{COSINUS}^\text{T}(E_\text{R}) \, \text{d}E_\text{R} \,.
\end{align}
These two equations can be combined to give
\begin{equation}
\sum_\text{T} \int \frac{\text{d}\bar{R}^\text{T}}{\text{d}E_\text{R}} \frac{\epsilon_\text{DAMA}^\text{T}(E_\text{R})}{S_\text{DAMA}^\text{bound}} \, \text{d}E_\text{R} \geq \sum_\text{T} \int \frac{\text{d}\bar{R}^\text{T}}{\text{d}E_\text{R}} \frac{\epsilon_\text{COSINUS}^\text{T}(E_\text{R})}{R_\text{COSINUS}^\text{bound}} \, \text{d}E_\text{R} \; .
\label{eq:ineq}
\end{equation}

Crucially, the mean differential event rate $\mathrm{d}\bar{R}^\text{T}/\mathrm{d}E_\mathrm{R}$ is the same for both DAMA and COSINUS, irrespective of the type of interaction, because they employ exactly the same detector material. This means that eq.~(\ref{eq:ineq}) can only hold if $\epsilon_\text{DAMA}^\text{T} (E_\text{R}) / S_\text{DAMA}^\text{bound} > \epsilon_\text{COSINUS}^\text{T}(E_\text{R})/R_\text{COSINUS}^\text{bound}$ for at least one target nucleus T and some value of $E_\text{R}$. Consequently, the COSINUS measurement excludes an explanation of the DAMA signal in terms of DM-nucleus scattering for \emph{arbitrary differential event rates} at more than 95\% CL if
\begin{equation}
 \dfrac{\epsilon_\text{COSINUS}^\text{T}(E_\mathrm{R})}{R_\text{COSINUS}^\text{bound}} > \dfrac{\epsilon_\text{DAMA}^\text{T}(E_\mathrm{R})}{S_\text{DAMA}^\text{bound}} \quad \text{ for all } E_\mathrm{R} \text{ and all } \text{T}\; .
 \label{eq:effcomparison} 
\end{equation}

\begin{figure}
\centering
\hspace*{-1.0cm}
\includegraphics[scale=0.91]{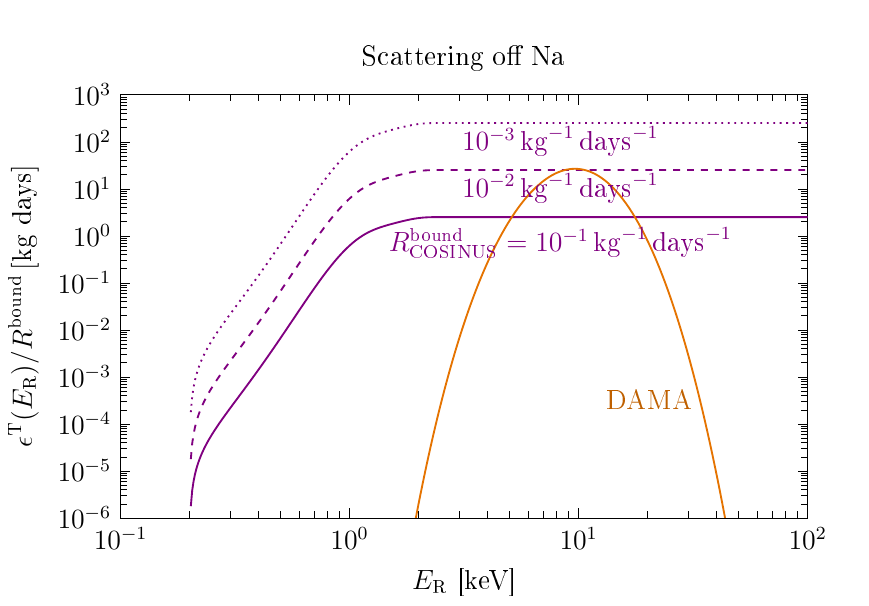}
\includegraphics[scale=0.91]{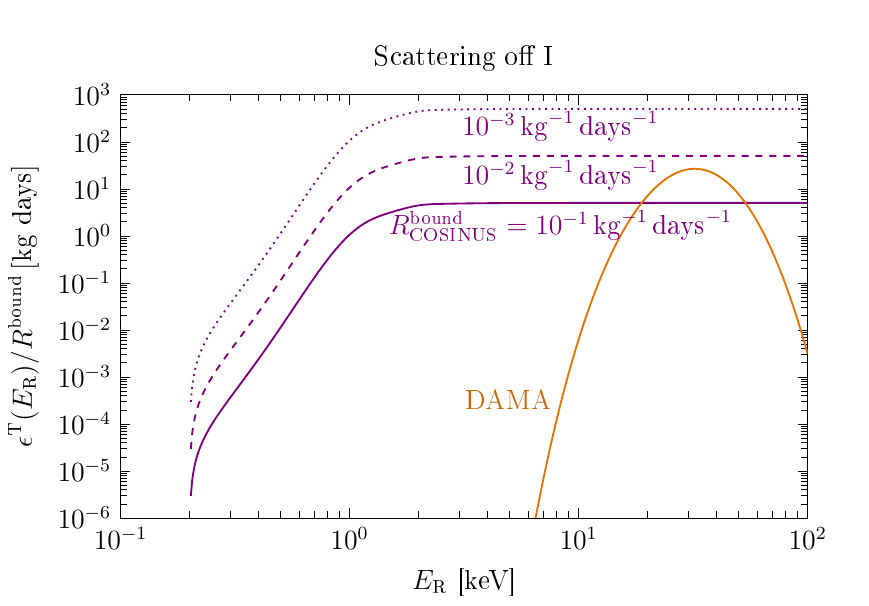}
\caption{\small Exposure-weighted efficiencies for DAMA (orange) and COSINUS (violet), corresponding to the two sides of eq.~(\ref{eq:effcomparison}). The different violet lines reflect different assumptions on $R_\text{COSINUS}^\text{bound}$, i.e.\ the bound on the total rate that COSINUS can achieve. If $R_\text{COSINUS}^\text{bound}$ is sufficiently small such that the COSINUS line lies above the DAMA line for all values of $E_\mathrm{R}$, the DAMA signal will be excluded for arbitrary differential event rates. This is the case for $R_\text{COSINUS}^\text{bound} < 10^{-2}\,\mathrm{kg^{-1}\,days^{-1}}$ for scattering off sodium and $R_\text{COSINUS}^\text{bound} < 2 \cdot 10^{-2}\,\mathrm{kg^{-1}\,days^{-1}}$ for scattering off iodine.}
\label{fig:comparison}
\end{figure}

In figure~\ref{fig:comparison} we show a comparison of the left-hand side and right-hand side of eq.~(\ref{eq:effcomparison}) for different assumptions on the value of $R_\text{COSINUS}^\text{bound}$ that can be achieved. The left panel corresponds to scattering off sodium, the right to scattering off iodine. As expected, the DAMA sensitivity peaks around $E_\mathrm{R} \approx 10\,\mathrm{keV}$ for sodium and around $E_\mathrm{R} \approx 33 \, \mathrm{keV}$ for iodine, both of which correspond to $E'_\mathrm{R} \approx 3\,\mathrm{keVee}$ when taking into account the appropriate quenching factors. The sensitivity of COSINUS, on the other hand, covers a much wider range of energies and in particular extends to significantly lower values of $E_\mathrm{R}$. In order to exclude the DAMA signal for sodium scatters, COSINUS thus needs to place a bound on the total event rate of $R_\text{COSINUS}^\text{bound} \approx 10^{-2} \, \text{kg}^{-1}\,\text{days}^{-1}$. A slightly weaker bound is sufficient for the case of iodine scatters, because the corresponding efficiency function in COSINUS is slightly larger. Since it is however not known what fraction of the DAMA signal is due to iodine scatters, a fully model-independent exclusion is only possible if COSINUS can exclude the DAMA signal for the case of scattering exclusively on sodium.

Eq.~(\ref{eq:effcomparison}) constitutes the most general method for comparing DAMA and COSINUS (or more generally any experiment observing an annual modulation with another experiment constraining the total rate for the same detector material). This approach is valid as long as the signal in both experiments is due to nuclear recoils. In particular it applies to all elastic and inelastic DM-nucleus scattering processes and to arbitrary velocity distributions. The sensitivity that COSINUS must achieve is however quite ambitious, about an order of magnitude beyond what we assumed for figure~\ref{fig:standard}. We will therefore now explore whether stronger constraints can be obtained for more restrictive assumptions on the differential event rate.

\section{Halo-independent comparison for classes of cross sections}
\label{sec:classes}

Let us begin by considering a fixed DM mass $m_\chi$ and a differential cross section with fixed shape but free normalisation, which is controlled by a parameter $\sigma$.\footnote{For the moment we assume that the differential event rate is dominated by either scattering off sodium or scattering off iodine. Below we will generalise to the case where both elements contribute at comparable level.} For a given DM velocity distribution $f(\mathbf{v})$ in the rest frame of the Sun, we define $R^f_\text{COSINUS}(\sigma)$ and $S^f_\text{DAMA}(\sigma)$ as the predicted rate and modulation amplitude in COSINUS and DAMA, respectively. With these conventions we define
\begin{equation}
 S_\text{DAMA}^\text{max} \equiv \max_f \big[ S_\text{DAMA}^{f}(\sigma) \text{ with } \sigma \text{ given by } R_\text{COSINUS}^{f} (\sigma)  = R_\text{COSINUS}^{\text{bound}}\big] \; .
  \label{eq:Smax}
\end{equation}
In other words, $S_\text{DAMA}^\text{max}$ is the largest modulation amplitude compatible with the COSINUS bound that can be obtained in DAMA from the assumed cross section for \emph{any DM velocity distribution}. If $S_\text{DAMA}^\text{max} < S_\text{DAMA}^\text{bound}$, we can therefore conclude that COSINUS excludes DAMA for the assumed cross section in a halo-independent way.

At first sight, calculating $S_\text{DAMA}^\text{max}$ seems rather challenging, given that it requires varying the three-dimensional function $f(\mathbf{v})$. To simplify the problem, we make the important observation that eq.~(\ref{eq:dRdE}) is linear in $f(\mathbf{v})$, meaning that we can write
\begin{align}
 R^f_\text{COSINUS}(\sigma) & = \int \mathrm{d}^3 \, v_0 \, f(\mathbf{v}_0) \, \mathcal{R}_\text{COSINUS}(\mathbf{v}_0, \sigma) \; .
 \label{eq:velocityexpansion}
\end{align}
Here $\mathcal{R}_\text{COSINUS}(\mathbf{v}_0, \sigma)$ is the rate calculated from the assumed differential cross section for a velocity distribution of the form $f(\mathbf{v}) = \delta^3(\mathbf{v} - \mathbf{v}_0)$, corresponding to a DM stream with a single velocity $\mathbf{v}_0$~\cite{Feldstein:2014gza}. In a fully analogous way, $S^f_\text{DAMA}$ can be expressed as a superposition of functions $\mathcal{S}_\text{DAMA}(\mathbf{v}_0, \sigma)$.

As shown explicitly in appendix~\ref{app:proof}, it turns out that in order to calculate $S_\text{DAMA}^\text{max}$, it is sufficient to consider velocity distributions of the form $f(\mathbf{v}) = \delta^3(\mathbf{v} - \mathbf{v}_0)$. In other words, the optimum velocity distribution that maximises $S_\text{DAMA}^f$ always corresponds to a single DM stream:\footnote{Note that for very small DM masses the optimal velocity can be significantly larger than the values typically assumed for the Galactic escape velocity $v_\text{esc}$. Restricting velocities to be smaller than $v_\text{esc}$ would lead to even stronger bounds.}
\begin{equation}
 S_\text{DAMA}^\text{max} \equiv \max_{\mathbf{v}_0} \big[ \mathcal{S}_\text{DAMA}(\mathbf{v}_0, \sigma) \text{ with } \sigma \text{ given by } \mathcal{R}_\text{COSINUS}(\mathbf{v}_0, \sigma) = R_\text{COSINUS}^{\text{bound}}\big] \; .
\end{equation}
At this point it is essential that we consider only a single energy range for both DAMA and COSINUS. The conclusion that a single DM stream is optimal no longer holds for experiments with several signal regions or experiments performing an unbinned analysis~\cite{Feldstein:2014ufa,Gondolo:2017jro}. In fact, the problem can be simplified further, because $\mathcal{R}_\text{COSINUS}(\mathbf{v}_0, \sigma)$ depends only on $v_0 = |\mathbf{v}_0|$, while $\mathcal{S}_\text{DAMA}(\mathbf{v}_0, \sigma)$ is maximized for streams that are anti-aligned with the velocity of the Earth on the 1st of June, given by $\mathbf{v}_\mathrm{E} \equiv v_\mathrm{E} \, \mathbf{e}_\text{summer}$ with $v_\text{E} = 30\,\text{km}\,\text{s}^{-1}$. It is therefore sufficient to consider only streams with $\mathbf{v}_0 = v_0 \, \mathbf{e}_\text{summer}$.\footnote{We note that gravitational focusing may modify the phase of the modulation signal~\cite{Lee:2013wza}, such that streams of a slightly different direction are needed to give the best fit to the DAMA signal.}

We conclude that, for a given cross section, COSINUS excludes DAMA in a halo-independent way if
\begin{equation}
 \frac{\mathcal{S}_\text{DAMA}(v_0, \sigma)}{S_\text{DAMA}^\text{bound}} <  \frac{\mathcal{R}_\text{COSINUS}(v_0, \sigma)}{R_\text{COSINUS}^\text{bound}}
 \label{eq:halo-ind}
\end{equation}
for all $v_0$. Note that, since the parameter $\sigma$ leads to the same rescaling on both sides, it is sufficient to check eq.~(\ref{eq:halo-ind}) for one specific value of $\sigma$. The quantities $\mathcal{R}_\text{COSINUS}$ and $\mathcal{S}_\text{DAMA}$ are easily calculated, allowing to determine very quickly whether DAMA and COSINUS can be compatible for a specific particle physics hypothesis. Alternatively, eq.~(\ref{eq:halo-ind}) can be rearranged to
\begin{equation}
R_\text{COSINUS}^\text{bound} < \min_{v_0} \left[ \frac{\mathcal{R}_\text{COSINUS}(v_0)}{\mathcal{S}_\text{DAMA}(v_0)} \right] S_\text{DAMA}^\text{bound} \; ,
\label{eq:power}
\end{equation}
i.e.\ we can calculate the bound on the total rate that COSINUS must achieve in order to exclude a specific hypothesis in a halo-independent way. We shall refer to this number as the \emph{exclusion power} of COSINUS.

The aim of this section is however to compare COSINUS and DAMA for whole classes of cross sections at once. The requirement derived above is insufficient for this purpose, as $\mathcal{R}_\text{COSINUS}$ and $\mathcal{S}_\text{DAMA}$ need to be recalculated as soon as the form of the differential cross section changes. However, we can make use of the same trick used to consider arbitrary velocity distributions to also consider broad classes of cross sections. Let us therefore assume that the differential cross sections of interest can be written as a linear combination of a set of basis functions
\begin{equation}
 \frac{\mathrm{d}\sigma^\mathrm{T}}{\mathrm{d}E_\mathrm{R}} = \sum_i \alpha^\mathrm{T}_i \frac{\mathrm{d}\sigma_i^\text{T}}{\mathrm{d}E_\mathrm{R}} \; ,
\end{equation}
where $\mathrm{T} = \text{Na},\text{I}$ denotes the two different target elements and we require $\alpha^\mathrm{T}_i \geq 0$. For each of these basis functions we can then calculate the total rate in COSINUS and the modulation amplitude in DAMA, called $\mathcal{R}^\mathrm{T}_{i,\text{COSINUS}}$ and $\mathcal{S}^\mathrm{T}_{i,\text{DAMA}}$, respectively. Note in particular that this calculation needs to be done separately for the two target elements. Following the same reasoning as above, one can immediately show that COSINUS excludes DAMA for \emph{any cross section of the assumed form} and \emph{any DM velocity distribution} if
\begin{equation}
 \frac{\mathcal{S}^\mathrm{T}_{i,\text{DAMA}}(v_0)}{S_\text{DAMA}^\text{bound}} <  \frac{\mathcal{R}^\mathrm{T}_{i,\text{COSINUS}}(v_0)}{R_\text{COSINUS}^\text{bound}}
\end{equation}
for all $v_0$, $i$ and $\mathrm{T}$. Clearly, this approach requires more calculational effort, but potentially allows to make very general statements.

\subsection{Monotonically decreasing cross sections}

Let us now turn to a particularly interesting example and consider the class of models for which the differential cross section is a monotonically decreasing function of recoil energy. More specifically, we assume that the differential cross section can be written as
\begin{equation}
\frac{\text{d}\sigma^\mathrm{T}}{\text{d}E_\mathrm{R}} = \frac{\sigma_0}{m_\text{T} v^2} \, \kappa^\mathrm{T}(E_\mathrm{R}) \, \Theta(E^\mathrm{T}_\text{max}(v) - E_\mathrm{R})
\label{eq:dsigmaTdER_monotonic} 
\end{equation}
with an arbitrary reference cross section $\sigma_0$ and dimensionless monotonically decreasing functions $\kappa^\mathrm{T}(E_\mathrm{R})$. In this expression the factor $1/v^2$ corresponds to the velocity dependence obtained in the usual case that the matrix element for the scattering does not depend on velocity and the final factor (with $E_{\text{max}}^\text{T}(v)  = 2 \mu_\mathrm{T}^2 v^2/m_\mathrm{T}$) enforces energy and momentum conservation for elastic scattering. Examples for cross sections that can be written in this way are spin-independent and spin-dependent scattering (see section~\ref{sec:specific}).\footnote{Strictly speaking, the relevant form factors only decrease monotonically for sufficiently small momentum transfer, but this covers the energy range relevant for COSINUS and DAMA.} Now we make use of the fact that any monotonically decreasing function can be written as a (potentially infinite) sum of step functions:
\begin{align}
\kappa^\mathrm{T}(E_\mathrm{R}) = \int_0^\infty \alpha^\mathrm{T}(E_0) \, \Theta(E_0 - E_\mathrm{R}) \mathrm{d}E_{0} \; .
\end{align}
We can therefore write any differential cross section of the form given in eq.~(\ref{eq:dsigmaTdER_monotonic}) as
\begin{equation}
\frac{\text{d}\sigma^\mathrm{T}}{\text{d}E_\mathrm{R}} = \int_0^\infty \alpha^\mathrm{T}(E_0) \frac{\text{d}\sigma^\mathrm{T}}{\text{d}E_\mathrm{R}}(E_0) \, \mathrm{d}E_{0}
\label{eq:expansion_decreasing}
\end{equation}
in terms of the basis functions
\begin{equation}
\frac{\text{d}\sigma^\mathrm{T}}{\text{d}E_\mathrm{R}}(E_0) = \frac{\sigma_0}{m_\text{T} v^2} \, \Theta(E_0 - E_\mathrm{R}) \, \Theta(E^\mathrm{T}_\text{max}(v) - E_\mathrm{R}) \; .
\label{eq:basisfcts_decreasing}
\end{equation}
Here the continuous parameter $\alpha^\mathrm{T}(E_0)$ takes the place of the discrete parameter $\alpha^\mathrm{T}_i$ introduced above.

In complete analogy to the previous results, we conclude that COSINUS excludes DAMA for \emph{any monotonically decreasing differential cross section} and \emph{any DM velocity distribution} if
\begin{equation}
 \frac{\mathcal{S}^\mathrm{T}_{\text{DAMA}}(v_0, E_0)}{S_\text{DAMA}^\text{bound}} < \frac{\mathcal{R}^\mathrm{T}_{\text{COSINUS}}(v_0, E_0)}{R_\text{COSINUS}^\text{bound}}
 \label{eq:halo-cs-ind}
\end{equation}
for all $v_0$, $E_0$ and $\mathrm{T}$. The great virtue of this expression is that $\mathcal{R}^\mathrm{T}(v_0, E_0)$ and $\mathcal{S}^\mathrm{T}(v_0, E_0)$ are in fact very easily calculated. Resubstituting the various definitions, we obtain the quite simple expressions
\begin{equation}
\mathcal{R}^\mathrm{T}(v_0, E_0) = \frac{\xi_\text{T} \, \rho \, \sigma_0}{m_\text{T}^2 \, m_\chi \, v_0} \int \epsilon^\mathrm{T}(E_\mathrm{R}) \, \Theta(E_0 - E_\mathrm{R}) \, \Theta(v_0 - v^\mathrm{T}_\text{min}(E_\mathrm{R})) \, \mathrm{d}E_\mathrm{R}
\label{eq:Rfullexpression}
\end{equation}
and
\begin{equation}
 \mathcal{S}^\mathrm{T}(v_0, E_0) \equiv \frac{1}{2} \left[ \mathcal{R}^\mathrm{T}(v_0^+, E_0) - \mathcal{R}^\mathrm{T}(v_0^-, E_0) \right]
\end{equation}
with
\begin{equation}
 v_0^+ = v_0 + v_\mathrm{E} \, , \qquad v_0^- = |v_0 - v_\mathrm{E}| \; .
\end{equation}
Note that $\mathcal{R}^\mathrm{T}(v_0, E_0)$ and $\mathcal{S}^\mathrm{T}(v_0, E_0)$ depend on the DM mass via $v^\mathrm{T}_\text{min}(E_\mathrm{R})$ and hence eq.~(\ref{eq:halo-cs-ind}) must be checked separately for each value of $m_\chi$. The local DM density $\rho$ as well as the reference cross section $\sigma_0$, on the other hand, cancel out in the condition~(\ref{eq:halo-cs-ind}).

\begin{figure}
\centering
\hspace*{-1.0cm}
\includegraphics[scale=0.91]{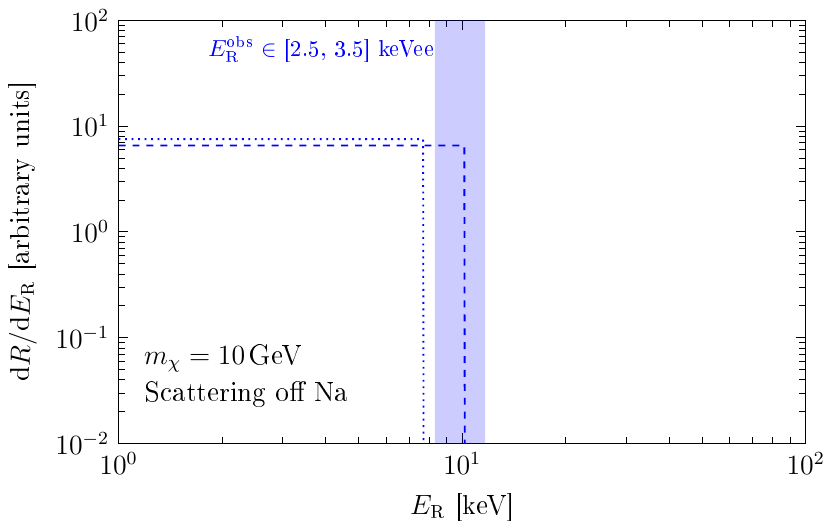}
\includegraphics[scale=0.91]{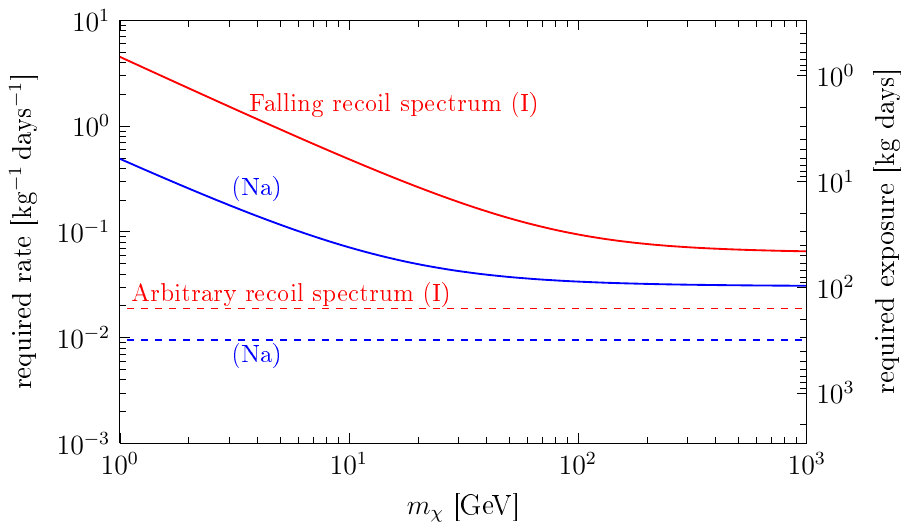}
\caption{\small Left: Differential recoil rate for scattering off sodium for the optimum monotonically decreasing scattering cross section, i.e.\ the scattering cross section satisfying eq.~(\ref{eq:dsigmaTdER_monotonic}) that maximises the modulation amplitude in DAMA while satisfying the (assumed) bound from COSINUS. The dashed (dotted) line corresponds to the scattering in summer (winter). Note that we have not included energy resolution and detector acceptance. Right: the exclusion power of COSINUS as defined in eq.~(\ref{eq:power}) for the case of monotonically decreasing scattering cross sections. The blue (red) line corresponds to scattering off sodium (iodine).}
\label{fig:falling}
\end{figure}

Clearly, for many values of $v_0$ and $E_0$ the modulation amplitude in DAMA will be zero or extremely small, either because the corresponding recoil energies are too small to be observable or because they are so large that DAMA is predicted to observe essentially the same rate in summer and winter. A large modulation fraction is obtained only if $v_0$ and $E_0$ are such that typical nuclear recoils are mostly below threshold in winter but slightly above threshold in summer. The configuration that maximises the modulation amplitude in DAMA relative to the total rate in COSINUS (for scattering off sodium and a DM mass of $m_\chi = 10\,$GeV) is illustrated in the left panel of figure~\ref{fig:falling}. The dashed (dotted) line indicates the differential event rate in summer (winter). Note that for small recoil energies the rate in winter is larger than in summer by a factor $v_0^+ / v_0^-$, see eq.~(\ref{eq:Rfullexpression}).

The blue shaded region in the left panel of figure~\ref{fig:falling} corresponds to the energy window probed by DAMA in the absence of fluctuations. For perfect energy resolution, one would therefore expect a modulation fraction in DAMA of $100\%$. The fact that the energy resolution in DAMA is however not perfect (and indeed pretty poor) means that some fraction of events will fluctuate into the signal region even in winter, so that the modulation fraction ends up being only about 46\%. To achieve larger modulation fractions, one would have to decrease $v_0$ further below the DAMA threshold. At this point it becomes crucial that COSINUS has a much lower threshold than DAMA, because reducing $v_0$ further would reduce the modulation amplitude in DAMA more rapidly than the total rate predicted for COSINUS, leading to a less than optimal configuration. 

The fact that the optimum choice of $v_0$ and $E_0$ for monotonically decreasing cross sections does not correspond to 100\% modulation fraction in DAMA means that we expect COSINUS to have more exclusion power than in the fully model-independent case discussed in section~\ref{sec:efficiencies}. In other words, weaker bounds on the total rate in COSINUS will be sufficient to exclude the DAMA signal. The values of $R_\text{COSINUS}^\text{bound}$ that COSINUS must achieve for this purpose are shown in the right panel of figure~\ref{fig:falling} as a function of $m_\chi$. Blue (red) solid lines correspond to scattering off sodium (iodine). The dashed lines show for comparison the values of $R_\text{COSINUS}^\text{bound}$ required for a fully model-independent exclusion (see figure~\ref{fig:comparison}).

We observe that the assumption of a monotonically decreasing cross section enhances the exclusion power of COSINUS by at least a factor of about three, meaning that three times weaker bounds on the rate (or three times less exposure) are sufficient to exclude the DAMA signal. For sufficiently small DM masses, the exclusion power becomes even greater owing to a reduced modulation fraction in DAMA. The nominal sensitivity of COSINUS $R_\text{COSINUS}^\text{bound} = 0.1 \mathrm{kg^{-1}\,days^{-1}}$ will be sufficient to exclude the DAMA signal for $m_\chi < 6 \,\text{GeV}$ for scattering off sodium and for $m_\chi < 90\,\mathrm{GeV}$ for scattering off iodine.

We emphasise that our approach is only suitable for addressing the question whether COSINUS can exclude the DAMA signal. If there is a combination of $v_0$ and $E_0$ such that
\begin{equation}
 \frac{\mathcal{S}^\mathrm{T}_{\text{DAMA}}(v_0, E_0)}{S_\text{DAMA}^\text{bound}} > \frac{\mathcal{R}^\mathrm{T}_{\text{COSINUS}}(v_0, E_0)}{R_\text{COSINUS}^\text{bound}}
\end{equation}
this does not necessarily imply that a good fit to the DAMA data can be obtained. The reason is that we consider only the modulation amplitude from DAMA in one specific bin and do not include the full information of the energy and time dependence of the DAMA signal.

\begin{figure}
\centering
\hspace*{-1.2cm}
\includegraphics[width=0.55\textwidth]{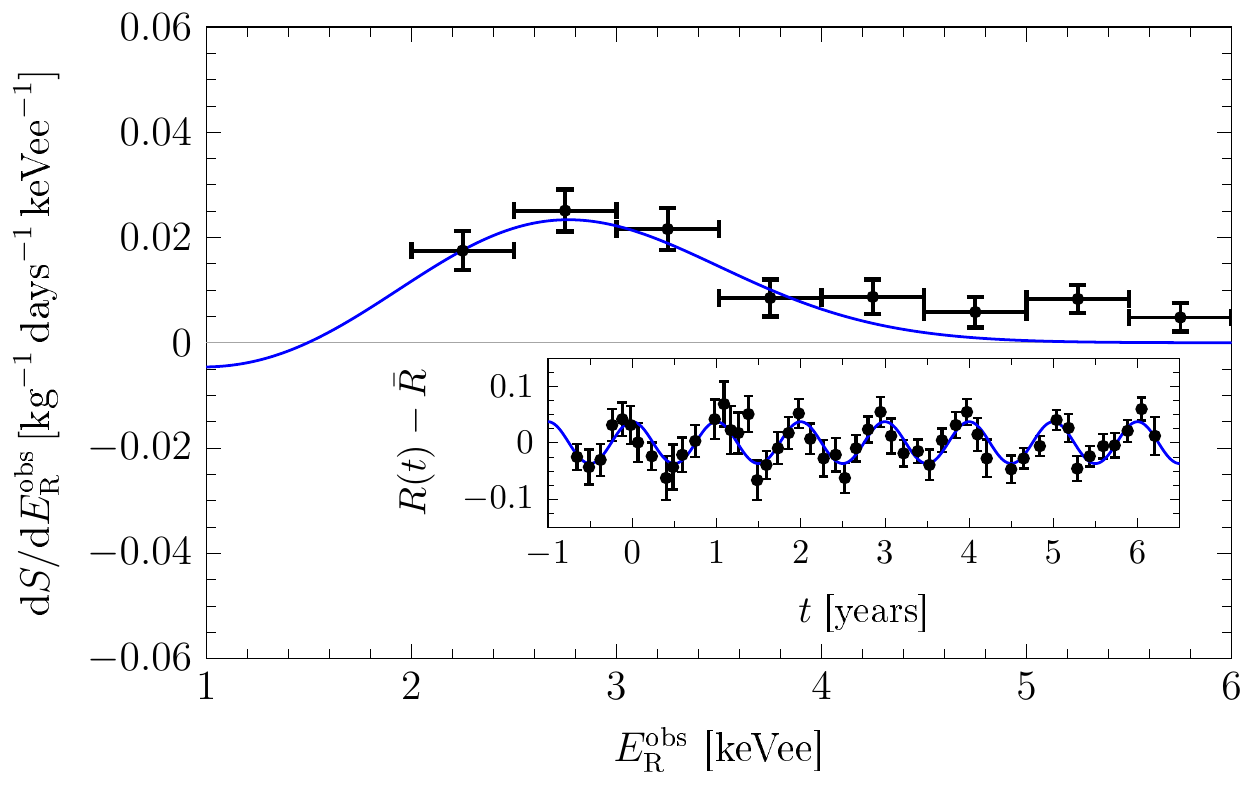}
\caption{\small Differential modulation amplitude $\mathrm{d}S/\mathrm{d}E_\text{R}^\text{obs}$ as a function of the \emph{observed} recoil energy $E_\text{R}^\text{obs}$ in DAMA for the optimum solution shown in figure~\ref{fig:falling}. The substantial difference between the two figures is due to the rather poor energy resolution of DAMA. Note that the data points in this plot (which are taken from the official DAMA analysis~\cite{Bernabei:2013xsa}) are shown only for illustration and have not been used to determine the optimum configuration, which is determined only from the total modulation amplitude in the bin $\left[2.5\,\text{keVee},3.5\,\text{keVee}\right]$. Nevertheless, both the energy dependence and the time dependence (see inset, which shows the energy-integrated modulation amplitude in units $\text{kg}^{-1}\,\text{days}^{-1}\,\text{keVee}^{-1}$) of the DAMA signal can be approximately reproduced, confirming that our treatment is neither too crude nor too conservative.}
\label{fig:SmearedSpectrum}
\end{figure}

Nevertheless, we make the interesting observation that the step function shown in the left panel of figure~\ref{fig:falling} actually gives a surprisingly good fit to the full DAMA data. The reason is once again that the poor energy resolution of the DAMA detector washes out the sharp feature in the differential event rate. This is illustrated in figure~\ref{fig:SmearedSpectrum}, which shows the differential modulation amplitude (with respect to observed energy $E_\mathrm{R}^\text{obs}$) as well as the time dependence of the modulation amplitude integrated over the energy interval $E_R^\text{obs} \in \left[2\,\text{keVee},6\,\text{keVee}\right]$ in comparison to the actual data points from DAMA. In fact, the quality of the fit is comparable to the one obtained for standard assumptions on the DM interactions and velocity distribution. This leads to the reassuring conclusion that the approach of decomposing the velocity distribution and the differential cross section into sets of basis functions is not exceedingly conservative.

\subsection{Momentum-dependent cross sections}

To conclude this section, let us consider possible alternatives to the assumption of monotonically decreasing cross sections. In principle, one could of course consider arbitrary cross sections, i.e.\ one could decompose the differential cross section into $\delta$-functions rather than step functions. It is easy to see, however, that in this case one can always arrange for 100\% modulation fraction in DAMA, so that one would recover the model-independent bounds obtained in section~\ref{sec:efficiencies}.

A more interesting assumption is that the differential cross section can grow no faster than some power $n$ of the recoil energy. In other words, we assume that the differential cross section can be written as
\begin{equation}
\frac{\text{d}\sigma^\mathrm{T}}{\text{d}E_\mathrm{R}} = \left(\frac{E_\mathrm{R}}{m_\mathrm{T}}\right)^n \frac{\sigma_0}{m_\text{T} v^2} \, \kappa^\mathrm{T}(E_\mathrm{R}) \, \Theta(E^\mathrm{T}_\text{max}(v) - E_\mathrm{R})
\label{eq:dsigmaTdER_momentum} 
\end{equation}
with monotonically decreasing functions $\kappa^\mathrm{T}(E_\mathrm{R})$. For $n = 0$ this corresponds to the case considered above, but for $n > 0$ one obtains less restrictive assumptions. As we will see in section~\ref{sec:specific}, this enables us to consider certain types of DM interactions that are not captured by the assumption of monotonically decreasing cross sections. The corresponding basis functions are given by
\begin{equation}
\frac{\text{d}\sigma^\mathrm{T}_n}{\text{d}E_\mathrm{R}}(E_0) = \left(\frac{E_\mathrm{R}}{m_\mathrm{T}}\right)^n \frac{\sigma_0}{m_\text{T} v^2} \, \Theta(E_0 - E_\mathrm{R}) \, \Theta(E^\mathrm{T}_\text{max}(v) - E_\mathrm{R}) \; .
\end{equation}

\begin{figure}
\centering
\hspace*{-1.2cm}
\includegraphics[width=0.55\textwidth]{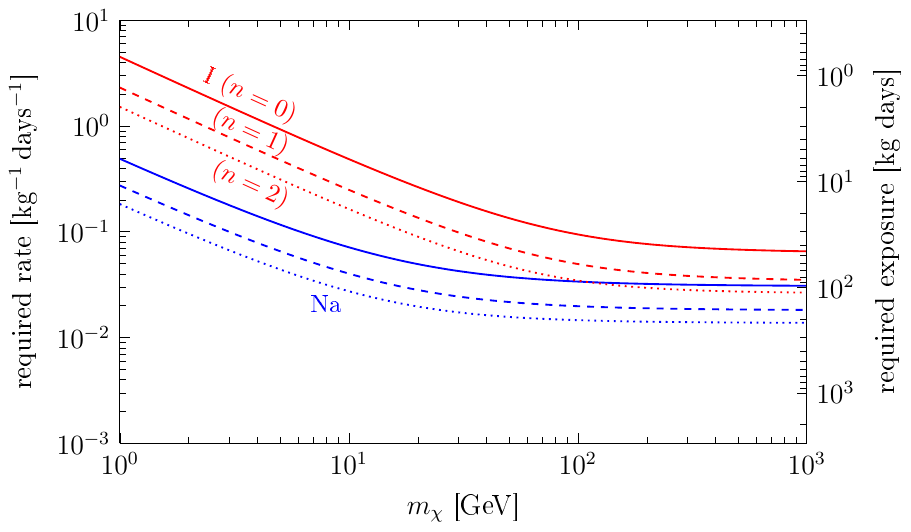}
\caption{\small Exclusion power of COSINUS for DM models in which the DM-nucleus scattering cross section can be written as a power $n$ of the recoil energy times a monotonically decreasing function (see eq.~(\ref{eq:dsigmaTdER_momentum})) for different values of $n$. Blue (red) lines correspond to scattering off sodium (iodine). The case $n = 0$ is identical to the case of a monotonically decreasing cross section considered in the right panel of figure~\ref{fig:falling}. For $n > 0$ the exclusion power of COSINUS is reduced due to the smaller benefit from the lower threshold.}
\label{fig:Comparison_Falling}
\end{figure}

The values of $R_\text{COSINUS}^\text{bound}$ required to exclude DAMA for different values of $n$ are shown in figure~\ref{fig:Comparison_Falling}. As expected, larger values of $n$, corresponding to less restrictive assumptions on the differential cross section, imply weaker exclusion power from COSINUS, meaning that stronger bounds on the rate (or larger exposure) are needed to exclude the DAMA signal. In fact, there is also a more intuitive way to understand figure~\ref{fig:Comparison_Falling}: As discussed above, much of the exclusion power of COSINUS comes from the fact that it has a lower threshold than DAMA. If the differential cross section grows with energy, the benefit of the lower threshold, and hence the exclusion power of COSINUS, is reduced.

\section{Halo-independent comparison for specific interactions}
\label{sec:specific}

In this section we compare the halo-independent bounds for different classes of recoil spectra derived above to the halo-independent bounds obtained for specific models that fall into these general categories. Specifically, we will consider spin-independent and spin-dependent scattering as examples for models with monotonically falling recoil spectra and $CP$-odd interactions as examples for models with momentum-dependent scattering. Towards the end of this section we will discuss inelastic DM as an example of a model that does not fall into one of the categories discussed in section~\ref{sec:classes}.

The crucial difference between the more general treatment above and the more specific treatment in this section is that by considering a specific type of interaction we impose certain relations between the scattering rates for different elements. Doing so implies in particular that DAMA faces overwhelming constraints from other direct detection experiments. There is consequently less motivation for performing a comparison between DAMA and COSINUS in the context of these models. Such a comparison is nevertheless instructive in order to explore how the general bounds derived above compare to the bounds obtained within specific models.

As we will see below, as long as the event rate in both DAMA and COSINUS is dominated by the scattering off the same target element, the general bounds derived above are only slightly weaker than the ones obtained within a specific model. However, much stronger exclusion limits are obtained in those regions of parameter space where scattering in DAMA is dominated by a single target element, while both sodium and iodine contribute to the event rate in COSINUS. In this case the performance of COSINUS turns out to strongly depend on its low-energy threshold.

\subsection{Spin-independent scattering}

We begin by discussing in detail the case of spin-independent scattering, which was also briefly considered in section~\ref{sec:DAMA} (see eq.~(\ref{eq:SI}) and figure~\ref{fig:standard}). Since the corresponding differential cross section is of the form of eq.~(\ref{eq:dsigmaTdER_monotonic}), we can immediately read off a conservative estimate of the exclusion power of COSINUS from figure~\ref{fig:falling}. A stronger bound can however be obtained by explicitly calculating $\mathcal{R}_\text{COSINUS}(v_0)$ and $\mathcal{S}_\text{DAMA}(v_0)$ for stream-like velocity distributions and making use of eq.~(\ref{eq:power}).

\begin{figure}
\centering
\hspace*{-1.0cm}
\includegraphics[scale=0.9]{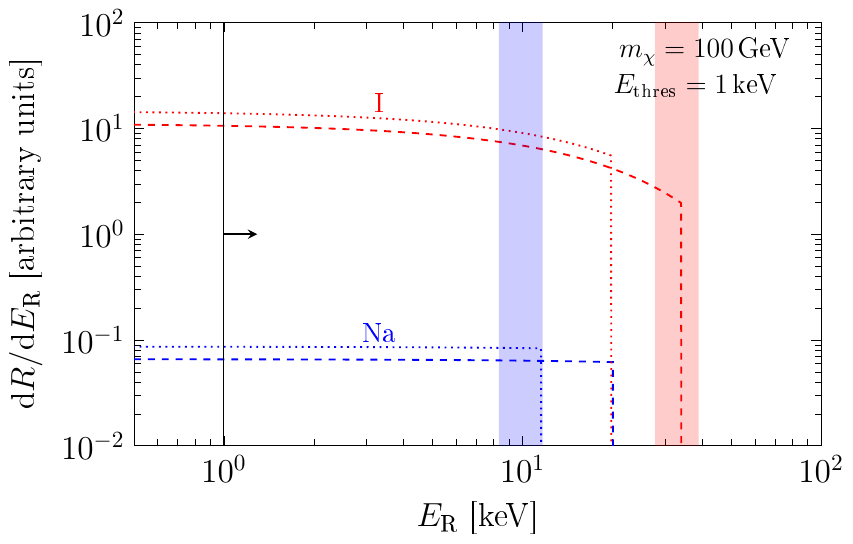}\hspace*{0.2cm}
\includegraphics[scale=0.9]{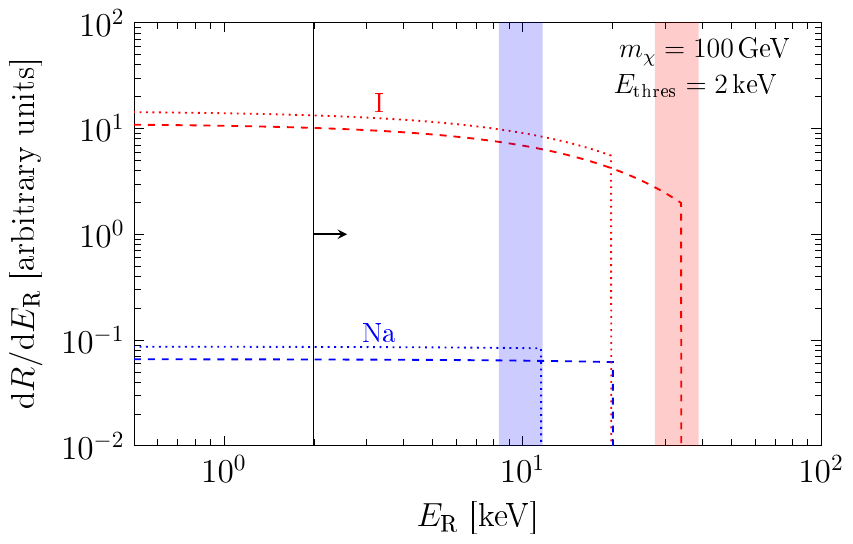}\\
\hspace*{-1.0cm}
\includegraphics[scale=0.9]{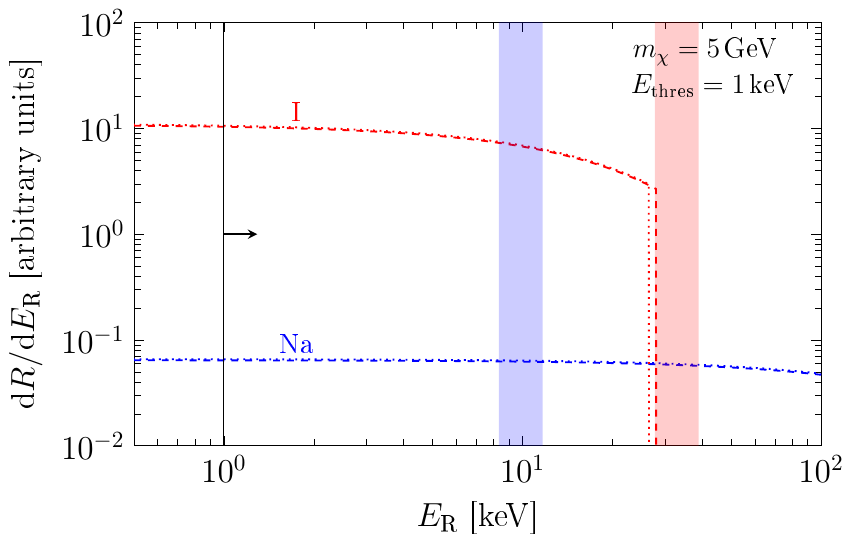}\hspace*{0.2cm}
\includegraphics[scale=0.9]{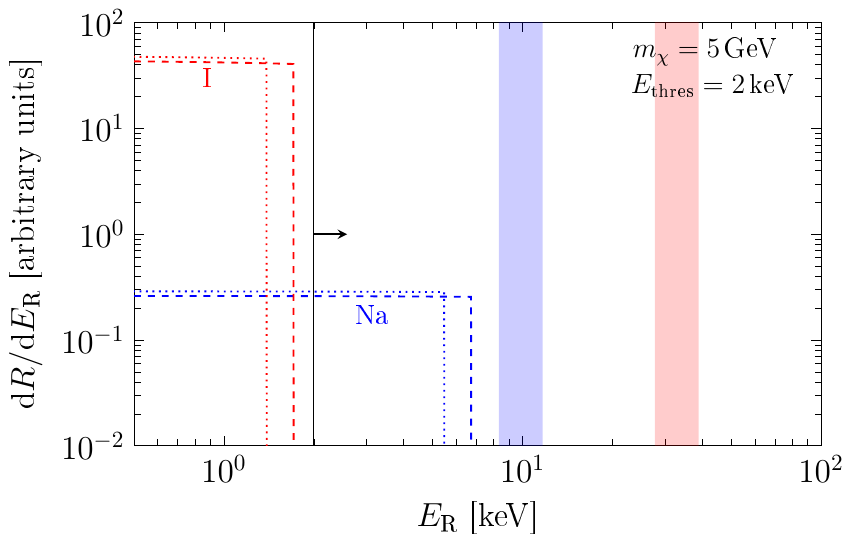}
\caption{\small Differential recoil rate (without energy resolution and detector acceptance) for spin-independent scattering with the optimum value of $v_0$. Blue (red) lines correspond to scattering off sodium (iodine), dashed (dotted) lines indicate the rates in summer (winter). The shaded regions indicate the approximate energy ranges that can contribute to the observed modulation in DAMA, the black arrow indicates the energy range that is relevant for COSINUS. The different rows (columns) correspond to different values of $m_\chi$~($E_\text{thres}$).}
\label{fig:dRdE_SI}
\end{figure}

Figure~\ref{fig:dRdE_SI} shows the recoil spectrum (in terms of the true nuclear recoil energy $E_\mathrm{R}$) for the optimal value of $v_0$, i.e.\ the one that imposes the strongest requirement on $R_\text{COSINUS}^\text{bound}$. The two different rows (columns) correspond to different values of $m_\chi$ ($E_\text{thres}$). As in the left panel of figure~\ref{fig:falling} the dashed (dotted) curves show the rates in summer (winter). The crucial difference to figure~\ref{fig:falling} is that for each case there are now two curves, corresponding to sodium (blue) and iodine (red).

We make a number of pertinent observations: First of all, we note that the rate from scattering off iodine is several orders of magnitude larger than the rate from scattering off sodium, owing to the well-known coherent enhancement for heavy target nuclei. Furthermore, as a result of the form factor suppression, the differential event rates are no longer constant below the cut-off. Finally, we note that for $m_\chi = 100\,\text{GeV}$ the differential event rate for iodine extends to larger recoil energies, while for $m_\chi = 5\,\text{GeV}$ the situation is reversed, due to the different ratios of DM particle mass to target nucleus mass.

For $m_\chi = 100\,\text{GeV}$ the situation is clear: Both the total rate and the modulation amplitude are completely dominated by scattering off iodine, and the contribution from sodium can effectively be neglected.\footnote{We note that scenarios in which the modulation amplitude in DAMA is dominated by scattering off iodine are already constrained by existing direct detection experiments based on target nuclei also involving iodine, such as COUPP~\cite{Behnke:2012ys}, KIMS~\cite{Lee:2014zsa} and PICO-60~\cite{Amole:2015pla}.} The only difference between spin-independent scattering and the general case considered in figure~\ref{fig:falling} is the form factor suppression, which substantially suppresses the event rate in the energy range probed by DAMA, but is less relevant for the lower energies probed by COSINUS. We therefore expect COSINUS to have somewhat higher exclusion power (by a factor of a few) for spin-independent interactions than for general monotonically decreasing scattering. Quantitatively, this is confirmed by comparing the orange and cyan curves in the lower left panel of figure \ref{fig:1}.

For $m_\chi = 5\,\text{GeV}$ the situation becomes more complicated, as the optimum value of $v_0$ depends on $E_\text{thres}$. The reason is that the value of $v_0$ that maximises the modulation fraction corresponds to scattering off sodium, such that scatters off iodine are completely below threshold for DAMA. However, this configuration is of advantage for DAMA only if iodine scatters are also below threshold in COSINUS. If the COSINUS threshold is low enough to be able to observe iodine scatters even if the DAMA signal is dominated by sodium scatters, the resulting bound from COSINUS will be overwhelmingly strong due to the coherent enhancement factor. Thus, for $E_\text{thres} = 1\,\text{keV}$ the optimum value of $v_0$ still corresponds to the case that scattering off iodine dominates both for DAMA and COSINUS. Scattering off sodium becomes advantageous for DAMA only if the threshold of COSINUS is rather high, starting at about $E_\text{thres} \simeq 2\,\text{keV}$.\footnote{We note that even for $m_\chi = 5\,\text{GeV}$ and $E_\text{thres} = 2\,\text{keV}$ the differential event rate due to scattering off iodine is only barely below threshold in COSINUS. As a result, the optimum configuration looks quite different from the one in figure~\ref{fig:falling}, with only upward fluctuations contributing to the DAMA signal. Only for even smaller DM masses and/or even higher thresholds the optimum configuration from figure~\ref{fig:falling} is recovered.}

\subsection{Spin-dependent and momentum-dependent scattering}

Having discussed in detail the case of spin-independent scattering, we now consider a number of alternative possibilities. In each case there is a well-defined relation between the scattering rate off iodine and the one off sodium, meaning that we expect COSINUS to have greater exclusion power than for model-independent assumptions. 

The elastic scattering of Galactic DM particles off nuclei can be fully characterised in terms of 18 non-relativistic effective operators, which can depend on the spin of the DM particle and the nucleon, $\mathbf{s}_\chi$ and $\mathbf{s}_N$, the relative velocity $\mathbf{v}$ and the momentum transfer $\mathbf{q}$~\cite{Fitzpatrick:2012ix}.\footnote{In addition, the cross section may be multiplied by an overall factor that depends on the momentum transfer in a target-independent way. For example, a factor of $q_\text{ref}^4 / q^4$ is present for the case of long-range interactions arising from the exchange of a very light mediator~\cite{Foot:2012rk}.} Within this nomenclature, standard spin-independent scattering corresponds to $\mathcal{O}_1$, while standard spin-dependent scattering corresponds to $\mathcal{O}_4$. Without reviewing the formalism in detail, we note that several of these operators predict cross sections that grow with increasing momentum transfer, corresponding to recoil spectra that do not decrease monotonically. Particularly well-known examples are the operators generated by the exchange of a spin-0 mediator with general $CP$ phase~\cite{Kahlhoefer:2017umn}. These are
\begin{align}
 \mathcal{O}_{10}^N & = i \, \mathbf{s}_N \cdot \mathbf{q}/m_N \,,\\
 \mathcal{O}_{11}^N & = i \, \mathbf{s}_\chi \cdot \mathbf{q}/m_N \,,\\
 \mathcal{O}_{6}^N & = (\mathbf{s}_N \cdot \mathbf{q})(\mathbf{s}_\chi \cdot \mathbf{q})/m_N^2 \,,
\end{align}
with $m_N$ the nucleon mass.
The differential event rate resulting from each of these operators can be written as some power $n$ of the nuclear recoil energy times a monotonically decreasing function (see eq.~(\ref{eq:dsigmaTdER_momentum})). Specifically, since $E_\mathrm{R} \propto q^2$, one finds $n = 1$ for $\mathcal{O}_{10,11}$ and $n = 2$ for $\mathcal{O}_6$.
In the following we assume that the corresponding spin-0 mediator couples to each Standard Model quark proportionally to its mass, as required by minimal flavour violation. Using the quark spin contents from ref.~\cite{Cheng:2012qr} and quark masses from ref.~\cite{Patrignani:2016xqp}, the neutron-to-proton coupling ratios of the operators are given by $\simeq 1$ for $\mathcal{O}_{11}^N$ and $-0.26$ for $\mathcal{O}_{6}^N$ and $\mathcal{O}_{10}^N$.

\begin{figure}
\centering
\hspace*{-0.7cm}
\includegraphics[scale=0.9]{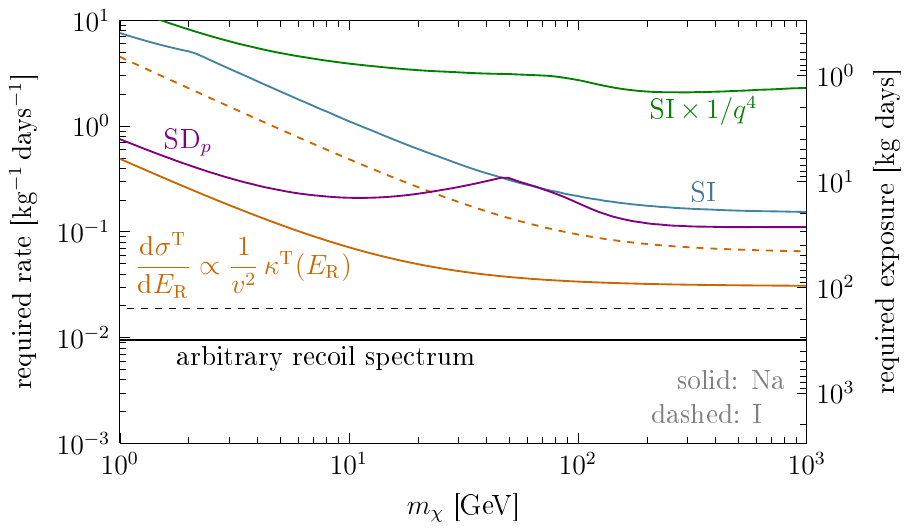}\hspace*{0.2cm}
\includegraphics[scale=0.9]{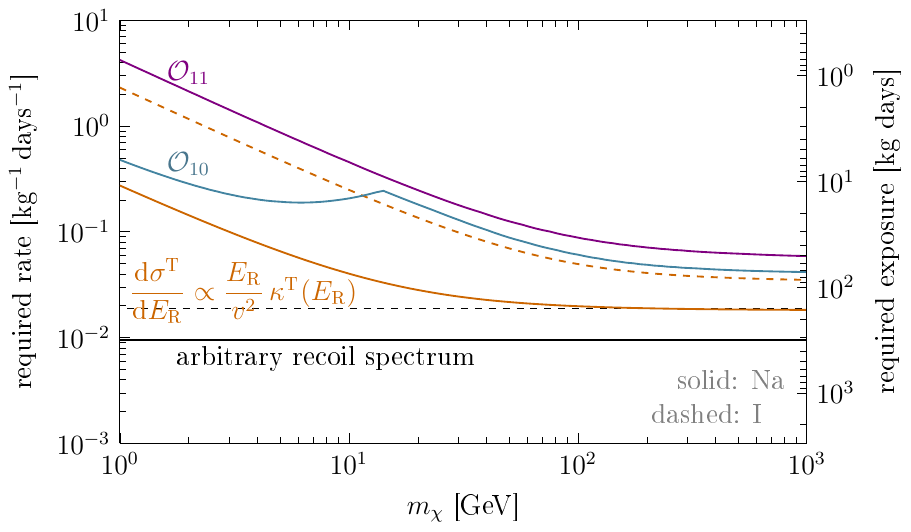}\\[0.2cm]
\hspace*{-0.7cm}
\includegraphics[scale=0.9]{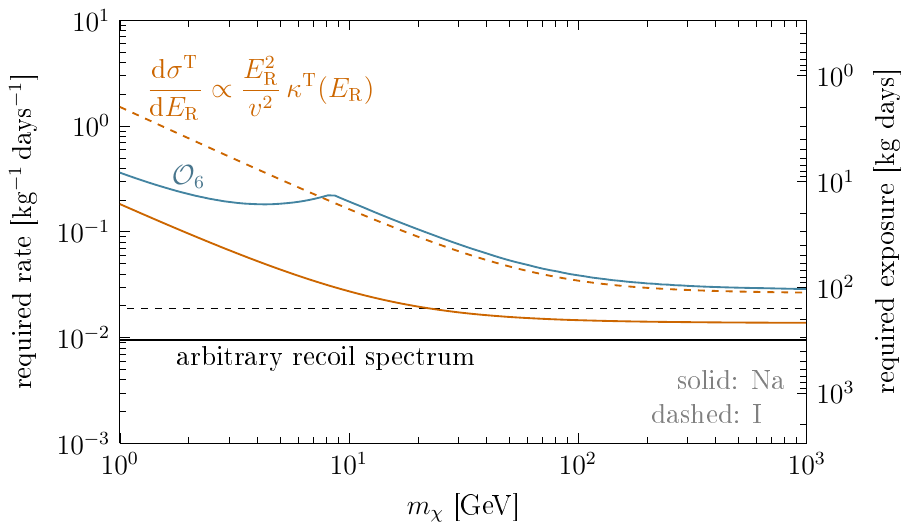}\hspace*{0.2cm}
\includegraphics[scale=0.9]{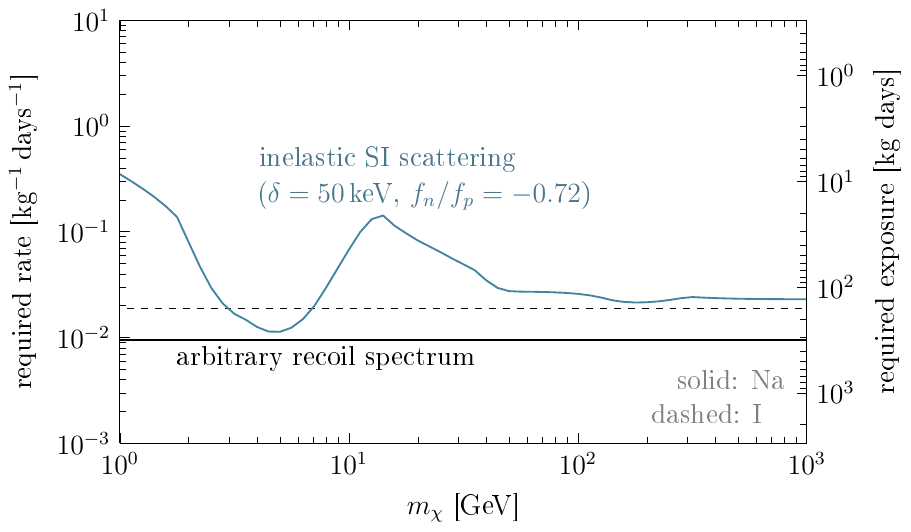}
\caption{\small COSINUS exclusion power for different assumptions regarding the underlying particle physics of DM-nucleus interactions. The first three panels consider interactions for which the cross section can be written as a power $n$ of the recoil energy times a monotonically decreasing function. In each panel the black lines show the fully model-independent constraint derived in section~\ref{sec:efficiencies} (solid for scattering off sodium, dashed for scattering off iodine), the orange lines show the bounds for classes of recoil spectra obtained from eq.~(\ref{eq:dsigmaTdER_momentum}), see also figure~\ref{fig:Comparison_Falling}, and the remaining lines correspond to specific interactions. The final panel considers the case of inelastic DM, which cannot be written in the form of eq.~(\ref{eq:dsigmaTdER_momentum}) and can therefore only be compared to the fully model-independent constraint.}
\label{fig:method_comparison}
\end{figure}

The COSINUS exclusion power for each of these operators can be calculated in complete analogy to the case of standard spin-independent scattering discussed above. On the one hand, this enables us to determine which particle physics explanations of the DAMA signal can be excluded by COSINUS in a halo-independent way. On the other hand, we can use the results to determine whether the more model-independent approach from section~\ref{sec:classes} is overly conservative. Our results are summarized in figure~\ref{fig:method_comparison}. The orange curves correspond to the ones already shown in figure~\ref{fig:Comparison_Falling} (solid for sodium, dashed for iodine), while the other curves correspond to specific particle physics assumptions (i.e.\ to specific non-relativistic operators). The first three panels correspond to $n=0,1,2$, the final panel will be discussed in the following sub-section.

In the first panel, we show in addition to spin-independent scattering also the case of spin-dependent scattering off protons (denoted SD$_p$). Due to the lack of coherent enhancement for scattering off iodine, the best-fit configuration for small DM masses corresponds to scattering dominantly off sodium. The transition from scattering off iodine to scattering off sodium leads to a visible kink in the exclusion power. We observe that in both regimes the exclusion power of COSINUS is quite close to the respective bounds derived in a model-independent way in section~\ref{sec:classes} (orange lines). Nevertheless, we conclude that the design sensitivity of COSINUS will be sufficient to exclude both spin-independent and spin-dependent scattering as an explanation of the DAMA signal in a halo-independent way, independently of the DM mass. We also show the case of spin-independent long-range interactions containing an additional factor of $1/q^4$ (denoted $\text{SI}  \times  1/q^4$). Due to the steeply falling recoil spectrum, the exclusion power of COSINUS is very large, and it should be easy for COSINUS to rule out any interpretation of DAMA in terms of long-range interactions (as proposed e.g.\ in ref.~\cite{Foot:2012rk}).

In the second and third panel we see that, as expected, an explanation of the DAMA signal in terms of momentum-suppressed scattering is more difficult to exclude with COSINUS, as its lower threshold is less beneficial. Nevertheless, for the design sensitivity $R_\text{COSINUS}^\text{bound} = 0.1\,\mathrm{kg^{-1}\,days^{-1}}$, COSINUS will be able to rule out also these scenarios in a halo-independent way for DM masses below $\simeq 20\,$GeV, $\simeq 40\,$GeV and $\simeq 75\,$GeV for scattering mediated by the non-relativistic operators $\mathcal{O}_{6}$, $\mathcal{O}_{10}$ and $\mathcal{O}_{11}$, respectively. The qualitative difference between the behaviour of $\mathcal{O}_{11}$ and the one of $\mathcal{O}_{6}$ and $\mathcal{O}_{10}$ is once again due to the absence of coherent enhancement for the latter two.

\subsection{Inelastic scattering}

It is well known that if DM-nucleus scattering is inelastic, i.e.\ if it requires the transition between two DM states of slightly different mass, the tension between the DAMA signal and other direct detection experiments is reduced~\cite{TuckerSmith:2001hy}. For an inelasticity parameter of $\delta$ (corresponding to the mass difference of the two DM states), the minimum velocity required for a recoil of energy $E_\mathrm{R}$ is given by
\begin{equation}
v_\text{min} = \left|\delta + \frac{m_\mathrm{T} \, E_\mathrm{R}}{\mu} \right| \frac{1}{\sqrt{2 \, E_\mathrm{R} \, m_\mathrm{T}}} \; .
\end{equation}
The requirement that $v_\text{min}$ is smaller than the Galactic escape velocity places a lower bound on $E_\mathrm{R}$, meaning that there are no low-energy nuclear recoils and as a result the modulation fraction can be large. 

The scattering rate for inelastic DM cannot be written in the form of eq.~(\ref{eq:dsigmaTdER_momentum}), providing a prime example for a scenario in which the rather general constraints derived in section~\ref{sec:classes} do not apply. Nevertheless, the fully model-independent bound derived in section~\ref{sec:efficiencies} does still apply, so we can immediately obtain a conservative estimate of the exclusion power of COSINUS. In the final panel of figure~\ref{fig:method_comparison} we study how this conservative estimate compares to the sensitivity that COSINUS must achieve to exclude a specific model of inelastic scattering as an explanation of the DAMA signal in a halo-independent way. For the latter, we choose for illustration the case of spin-independent scattering with $f_n/f_p \simeq -0.72$ and a mass splitting $\delta = 50\,$keV. Similar scenarios have been studied e.g.\ in refs.~\cite{Frandsen:2011ts,Scopel:2014kba} in order to suppress the scattering rate in xenon- and/or germanium-based experiments.\footnote{The idea of inelastic DM was recently revisited in ref.~\cite{Kang:2018dlc} for a scenario with proto-philic spin-dependent coupling to nucleons. We find that such a scenario can be probed by COSINUS with much less exposure, as in contrast to spin-independent interactions with $f_n/f_p \simeq -0.72$ there is no suppression for scatterings off iodine recoils.} 

The mass dependence of the rate required in COSINUS for excluding a DM interpretation of DAMA can be understood as follows. For $m_\chi \lesssim 5\,$GeV, the suppression of iodine scatters due to the choice $f_n/f_p \simeq -0.72$ leads to a situation in which both the rate in COSINUS and DAMA are dominated by sodium recoils. For these DM masses, the range of kinematically allowed energies corresponding to the optimal choice of the stream velocity $v_0$ gets more and more narrow with increasing $m_\chi$. At $m_\chi \simeq 5\,$GeV, the situation is already nearly equivalent to adopting a single recoil energy $E_\text{R} = E_0$ for scattering off sodium, and hence the fully model-independent bound shown as a solid black curve is almost saturated. For $5\,\text{GeV} \lesssim m_\chi \lesssim 10\,\text{GeV}$, iodine recoils start to become relevant in COSINUS but not in DAMA, due the larger threshold of the latter experiment; consequently, a less stringent bound on the rate in COSINUS is necessary for excluding DAMA. Finally, for $m_\chi \gtrsim 10\,$GeV iodine recoils dominate over sodium both in COSINUS and DAMA (see also the lower panel in figure~\ref{fig:dRdE_SI}). The initial decrease of the required rate for these DM masses is again due to the smaller range of kinematically allowed energies, favouring DAMA over COSINUS. For $m_\chi \gtrsim 50\,$GeV, the situation is again nearly identical to the case of recoils with a single energy, such that in this case the fully model-independent bound for scattering off iodine is nearly saturated.

\section{Conclusions}
\label{sec:conclusions}

While the interpretation of the DAMA annual modulation in terms of DM is strongly challenged by null results from other direct detection experiments, it is extremely difficult to refute this interpretation in a model-independent way. In particular, the different target elements as well as techniques employed by the various experiments, while offering important complementarity, also imply larger uncertainties due to the unknown local velocity distribution as well as the particle physics properties of DM. 

In this article we have discussed how the DM interpretation of the DAMA signal in terms of nuclear scatterings can be tested independently of astrophysical and particle physics unknowns with COSINUS, a future experiment which employs the same target material and is thus sensitive to exactly the same recoil spectrum as DAMA (before taking into account detection efficiencies). Compared to other planned experiments also based on a NaI target, a great advantage of the cryogenic detector developed by the COSINUS collaboration lies in its ability to measure the {\it total} DM induced scattering rate with a significantly lower threshold than DAMA.

Our main findings are summarised in figure~\ref{fig:1}, where we show the halo-independent exclusion power of COSINUS, i.e.\ the bound on the total rate that COSINUS must achieve for excluding DAMA in a halo-independent way, as a function of the assumed threshold in COSINUS. The vertical axis on the right of each panel indicates the required exposure assuming zero observed events. The four panels correspond to different values of the DM mass $m_\chi$, the various lines in each panel result from different assumptions on the differential event rate. We make the following observations:
\begin{enumerate}
 \item The black shaded regions in figure~\ref{fig:1} show which combinations of the bound on the total rate and the threshold in COSINUS suffice to exclude a DM interpretation of DAMA, \emph{fully independently} of the assumed nuclear recoil spectrum. This is only possible if the COSINUS threshold is significantly lower than the one of DAMA, $E_\text{thres} \lesssim 1.8\,$keV.  While this is compatible with the design goal $E_\text{thres} = 1\,\text{keV}$, the required bound on the total rate is about an order of magnitude stronger than the current sensitivity estimate $R_\text{COSINUS}^\text{bound} = 0.1 \mathrm{\,kg^{-1}\,days^{-1}}$ for COSINUS.
 \item More restrictive assumptions on the differential scattering cross section of DM lead to greater exclusion power, meaning that a weaker bound on the total rate (or equivalently a smaller exposure) is sufficient to exclude the DAMA signal. In particular, the orange curves in figure~\ref{fig:1} show the required bound for the still very general class of differential scattering cross sections that are monotonically decreasing in energy and have the standard $1/v^2$ dependence on velocity. For the design sensitivity of COSINUS, such scenarios can be robustly excluded for DM masses $m_\chi \lesssim 6\,\text{GeV}$. If the COSINUS sensitivity is further increased by a factor of $\simeq 3$, this class of scattering cross sections can be excluded for all relevant DM masses. Interestingly, it follows from figure~\ref{fig:1} that this conclusion would still hold if the threshold in COSINUS is raised to significantly larger values; this is a prime example of how our results can be employed in order to optimise the detector layout of COSINUS for testing the DAMA hypothesis.
 \item Finally, we show in figure~\ref{fig:1} the required bound on the rate in COSINUS when assuming a given particle physics scenario for the scattering process of DM with nuclei (in this case standard spin-independent interactions). In contrast to the other curves, this fixes the relation between the scattering rate off sodium and iodine. In particular, depending on the COSINUS threshold and the DM mass, it can happen that the modulation signal in DAMA is dominated by sodium recoils, while COSINUS can also observe the corresponding signal from iodine, thanks to its lower threshold. This further lowers the required sensitivity of COSINUS compared to e.g.~the more general results based on monotonically decreasing cross sections. Similar results for various other interaction types are shown in figure~\ref{fig:method_comparison}.
\end{enumerate}
In summary, we have shown that if COSINUS is able to set a bound on the scattering rate of DM of $\simeq (0.01 - 0.1)\,\text{kg}^{-1}\,\text{days}^{-1}$ with an energy threshold of $\lesssim 1.8\,$keV, it can exclude very general classes of interactions in which the signal is dominated by DM scattering off sodium or iodine in a halo-independent way. As other explanations of the DAMA data such as DM-electron scattering~\cite{Foot:2014xwa}, scattering off thallium~\cite{Chang:2010pr} or scattering off OH impurities~\cite{Profumo:2014mpa} are already strongly constrained by other experiments~\cite{Aprile:2015ade,Petricca:2017zdp,Agnese:2017jvy}, this would strongly point towards a so far unknown source of background being responsible for the modulation seen by the DAMA collaboration.

To conclude, let us mention that for the last few years DAMA has been taking data with improved photo-multiplier tubes, which promise a significant lowering of the threshold~\cite{Bernabei:2012zzb}. Measurements of the modulation amplitude in this new energy range will provide important clues to the origin of the DAMA signal~\cite{Kelso:2013gda}. To make full use this new information it will be crucial to extend the methods developed here to also include spectral information when calculating the model-independent exclusion power of COSINUS. Clearly, the fact that COSINUS can achieve an even lower threshold and a much better energy resolution than DAMA will then become particularly important.

\section*{Note added}

\begin{figure}
\centering
\hspace*{-0.7cm}
\includegraphics[width=0.45\textwidth]{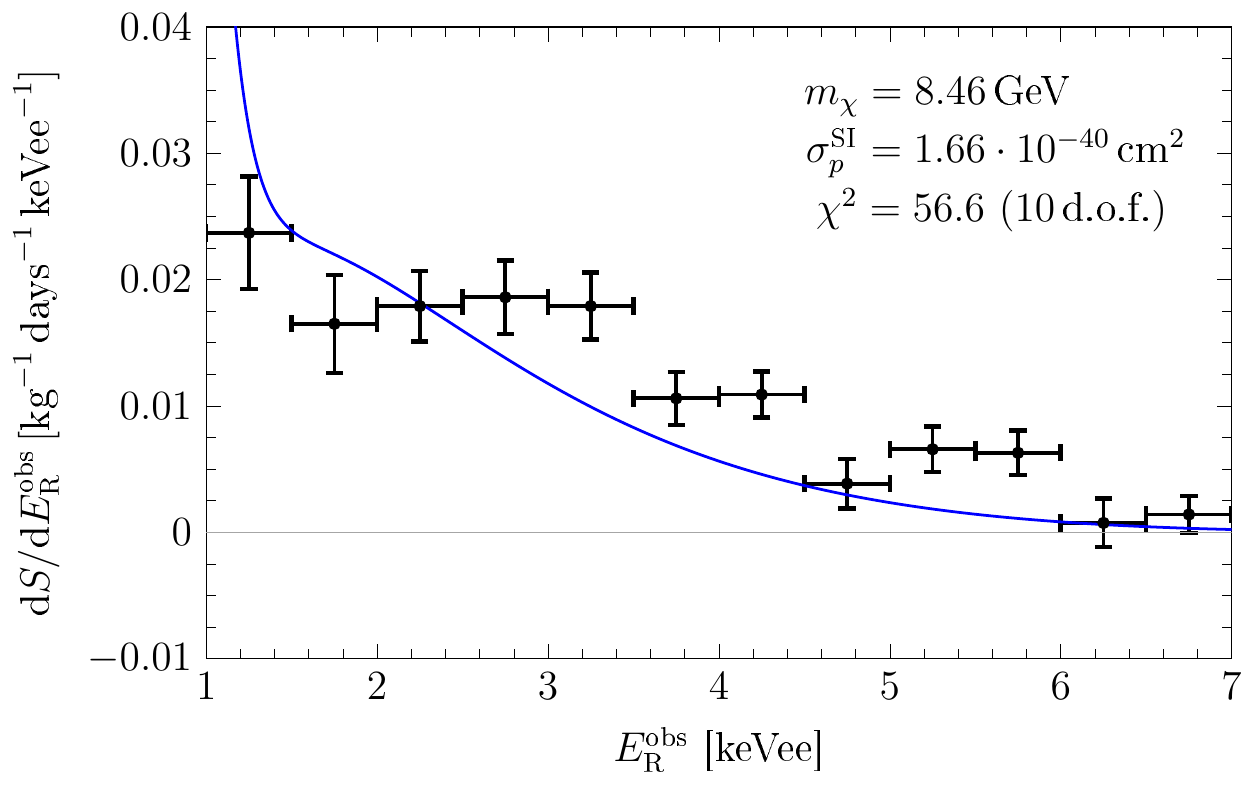}\hspace*{0.2cm}
\includegraphics[width=0.45\textwidth]{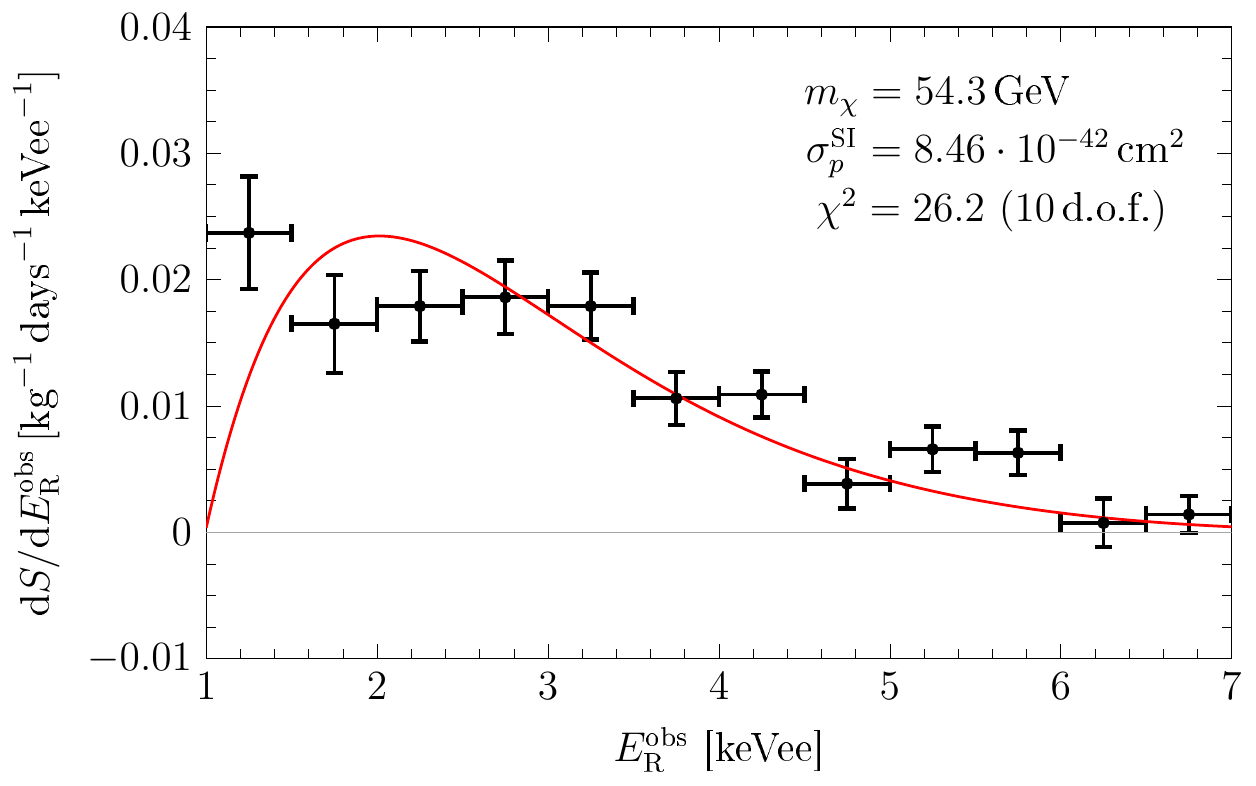}\\[0.2cm]
\caption{\small Best-fit recoil spectra in DAMA for low-mass DM (left), corresponding to scattering dominantly on sodium, and high-mass DM (right), corresponding to scattering dominantly on iodine. In both cases we have included the first twelve bins from the combined data sets of DAMA/NaI, DAMA/LIBRA-phase1 and DAMA/LIBRA-phase2.}
\label{fig:DAMAnew}
\end{figure}

After the completion of this work, DAMA presented new data with a lower threshold of $1\;\mathrm{keVee}$~\cite{DAMA_talk}. While the measurements above $2\,\mathrm{keVee}$ are in agreement with previous results, two new data points at lower energies provide very interesting additional information, because they exhibit neither the steep rise expected for scattering of low-mass DM, nor the turn-over towards anti-modulation expected for high-mass DM~\cite{Kelso:2013gda}. We have performed a combination of the two data sets, assuming that the energy resolution has not changed significantly during the detector upgrade. We find that it is no longer possible to obtain a good fit for DM scattering with the standard assumptions discussed in section~\ref{sec:DAMA}. The best-fit spectra for spin-independent scattering on sodium and spin-independent scattering on iodine are shown in figure~\ref{fig:DAMAnew}. The goodness of fit of the best-fit point for scattering on sodium is  unacceptable ($p \sim 10^{-8}$ when including the first 12 bins), while the hypothesis of scattering on iodine is only marginally compatible with data ($p \approx 0.003$). This observation implies that any interpretation of the DAMA signal in terms of DM requires non-standard interactions or non-standard astrophysical distributions (or both), independently of (but already implied by) the exclusion bounds from other experiments. New experiments based on NaI, like COSINUS, and model-independent methods, like the one presented in this work, will therefore be essential to further investigate the nature of the DAMA signal.

\acknowledgments

We thank Suchita Kulkarni, Josef Pradler and Thomas Schwetz for discussion and Vanessa Zema for helpful comments on the manuscript. This work is supported by the German Science Foundation (DFG) under the Collaborative Research Center (SFB) 676 ``Particles, Strings and the Early Universe'' and the Emmy Noether Grant No.\ KA 4662/1-1 as well as the ERC Starting Grant `NewAve' (638528). 

\appendix

\section{Best-fit velocity distribution}
\label{app:proof}

In this appendix we prove that a single stream is enough to determine $S_\text{DAMA}^\text{max}$ as defined in eq.~(\ref{eq:Smax}). First of all, we rewrite this equation as
\begin{align}
S_\text{DAMA}^\text{max} = \max_f \frac{S^f_\text{DAMA}(\sigma)}{R_\text{COSINUS}^{f}(\sigma) / R_\text{COSINUS}^\text{bound}} \; .
\label{eq:ratio}
\end{align}
The right-hand side of this equation is in fact independent of $\sigma$, as the cross section cancels out in the ratio. It is therefore sufficient to calculate $S_\text{DAMA}^\text{max}$ for an arbitrary reference cross section, so that we will simply drop the dependence on $\sigma$ for the rest of this appendix.

Let us now define
\begin{align}
S_\text{DAMA}^\text{stream} \equiv \max_{\mathbf{v}_0} \frac{S^{\mathbf{v}_0}_\text{DAMA}}{R_\text{COSINUS}^{\mathbf{v}_0} / R_\text{COSINUS}^\text{bound}}
\label{eq:ratio_stream}
\end{align}
in complete analogy to eq.~(\ref{eq:ratio}), but considering only DM velocity distributions of the form $f(\mathbf{v}) = \delta(\mathbf{v} - \mathbf{v}_0)$.  We now want to show that
\begin{align}
S_\text{DAMA}^\text{max} \equiv S_\text{DAMA}^\text{stream} \; .
\label{eq:claim}
\end{align}

Since it is obviously true that $S_\text{DAMA}^{\text{max}} \geq S_{\text{DAMA}}^\text{stream}$, it remains to be shown that $S_\text{DAMA}^{\text{max}} \leq S_{\text{DAMA}}^\text{stream}$. To do so, we denote the velocity that maximises the right-hand side of eq.~(\ref{eq:ratio_stream}) as $\mathbf{w}_0$, such that
\begin{align}
S_\text{DAMA}^\text{stream} = \frac{S^{\mathbf{w}_0}_\text{DAMA}}{R_\text{COSINUS}^{\mathbf{w}_0} / R_\text{COSINUS}^\text{bound}} \; .
\end{align}
For a given velocity distribution $f(\mathbf{v})$ we can then consider the following expression:
\begin{align}
 R_\text{COSINUS}^{f} \,\,-\,\, \frac{S_\text{DAMA}^{f}}{S_\text{DAMA}^{\mathbf{w}_0} / R_\text{COSINUS}^{\mathbf{w}_0}} = \int \text{d}^3 v_0 \, f(\mathbf{v}_0) \left(  R_\text{COSINUS}^{\mathbf{v}_0} \,\,-\,\, \frac{S_\text{DAMA}^{\mathbf{v}_0}}{S_\text{DAMA}^{\mathbf{w}_0} / R_\text{COSINUS}^{\mathbf{w}_0}} \right)
\label{eq:expansion_v0} \; ,
\end{align}
where we have made use of the fact that the velocity distribution enters linearly into the calculation of $R_\text{COSINUS}$ and $S_\text{DAMA}$ (see eq.~(\ref{eq:velocityexpansion})).
By definition of $\mathbf{w}_0$ we have
\begin{equation}
\frac{S_\text{DAMA}^{\mathbf{w}_0}}{R_\text{COSINUS}^{\mathbf{w}_0}} \, \geq \,\frac{S_\text{DAMA}^{\mathbf{v}_0}}{R_\text{COSINUS}^{\mathbf{v}_0}}
\end{equation}
for all $\mathbf{v}_0$, and hence the integrand in the right-hand side of eq.~(\ref{eq:expansion_v0}) is always positive. It therefore follows immediately that
\begin{equation}
R_\text{COSINUS}^{f} \,\,-\,\, \frac{S_\text{DAMA}^{f}}{S_\text{DAMA}^{\mathbf{w}_0} / R_\text{COSINUS}^{\mathbf{w}_0}} \geq 0 \; ,
\end{equation}
which can be rearranged to 
\begin{align}
\frac{ S_\text{DAMA}^{f}}{R_\text{COSINUS}^{f} / R_\text{COSINUS}^\text{bound}} \leq \frac{ S_\text{DAMA}^{\mathbf{w}_0}}{R_\text{COSINUS}^{\mathbf{w}_0} / R_\text{COSINUS}^\text{bound}} = S_\text{DAMA}^\text{stream} \; .
\end{align}
Since this inequality holds for arbitrary velocity distributions $f$, it must hold in particular for the velocity distribution that maximises the left-hand side, thus completing the proof.

\providecommand{\href}[2]{#2}\begingroup\raggedright\endgroup

\end{document}